\documentclass[aps,twocolumn,groupedaddress]{revtex4}
\usepackage[section]{placeins}        
\usepackage{float}

\usepackage{epsfig}
\usepackage{pdfpages}
\usepackage{amsmath,amsthm, amssymb, latexsym}
\usepackage{color}
\usepackage{graphicx}
\usepackage{epstopdf}
\usepackage{epsfig}
\usepackage[outercaption]{sidecap}

\newcommand{\bee}{\begin{equation}}
\newcommand{\ene}{\end{equation}}
\newcommand{\beea}{\begin{eqnarray}}
\newcommand{\enea}{\end{eqnarray}}

\begin{document}
\title{  Excitation of lower hybrid and magneto-sonic perturbations  in  laser plasma interaction }

\author{Ayushi Vashistha$^{1,2}$}
\thanks{ayushivashistha@gmail.com}
\author{Devshree Mandal$^{1,2}$}
\author{Amita Das$^{3}$}
\thanks{amitadas3@yahoo.com}

\affiliation{$^1$ Institute for Plasma Research, HBNI, Bhat, Gandhinagar - 382428, India } 
\affiliation{$^2$ {Homi Bhabha National Institute, Mumbai, 400094 } }

\affiliation{$^3$ Physics Department, Indian Institute of Technology Delhi,  Hauz Khas, New Delhi - 110016, India }

\begin{abstract} 
Lower hybrid (LH) and magneto-sonic  (MS) waves are  well known modes of  magnetized plasma. These modes play important roles in many phenomena. The lower hybrid wave  is often  employed  in magnetic confinement fusion experiments for  current drive and heating purposes.   Both LH and MS  waves  are  observed  in various astrophysical and  space plasma observations. These  waves involve ion motion and have not therefore been considered in high power pulsed laser experiments. 
  This paper shows, with the help of Particle - In - Cell (PIC) simulations,
a simple mechanism for excitation of lower hybrid and magnetosonic excitations  in the context of laser plasma interaction.  
A detailed study characterising the formation and  propagation of these modes have been provided. 
 The scheme for generating these perturbations  relies on the application of  a strong magnetic field in the plasma 
to constrain the motion of lighter electron species 
 in the laser electric field. The magnetic field strength is chosen 
 so as to leave the heavier ions  un-magnetized at the  laser frequency.  This helps in the excitation of the LH waves. 
On the other hand at the slower time scale associated with the laser pulse duration, even the ions show a magnetized response and magnetosonic excitations are observed to get excited.


\end{abstract}
\maketitle 

\section{Introduction}
The  magnetized plasma modes,  like lower hybrid (LH) and other modes 
(e.g. electron and ion cyclotron modes) have  found an important place in the context of magnetic confinement fusion studies.   They are  traditionally employed for the purpose of current drive and  heating of fusion plasma in magnetic confinement devices. Lower hybrid waves have been extensively studied in a variety of contexts in laboratory and space plasmas \cite{mcclements1997, bingham1997, fabio, southwood1978}. The LH waves can impart their energy to plasma species through various mechanisms like electron heating via breaking of lower hybrid wave \cite{fabio}, providing energy to electrons via wave-particle interaction. An application of the latter is found in toroidal tokamak devices where
landau damping of externally launched lower hybrid wave helps in driving plasma current required for plasma confinement \cite{lhcd, LHCD_NF}. Furthermore, it is very common to observe lower hybrid waves and several interesting phenomena due to their breaking or turbulence in space plasmas  \cite{retterer_ion_suprauroral, kelley1990_geophysical}. Localised lower hybrid turbulence is observed in density depletion regions in the early phases of an intense magnetic storm \cite{malingre2008lightning}. Many studies in the auroral region report observation of lower hybrid emissions, giving rise to solitary structures and/or ion-heating \cite{retterer_ion_suprauroral, berthelier2008lightning}. The LH mode  is unique in the sense that  the dynamical response of both  electron and  ions are   relevant for this mode.  Basically, the magnetized electron response couples with the un-magnetized ion dynamics for this particular mode to get excited. 

The linear modes for magnetized plasma  have, however, not been explored in the context of laser plasma interaction studies.  With  recent technological progress, quasi-static 
magnetic field of the order of $1.2$ kilo Tesla \cite{Nakamura} can   be produced  in  laboratory.   With rapid advancements in technology it is quite likely that the regime of magnetized plasma response at laser frequencies 
can be within reach of laboratory experiments. In the context of $CO_2$ lasers the magnetic field requirement to observe magnetized electron response 
(i.e. $\omega_{ce} > \omega_l$) turns out to be around $1.2$ kilo Tesla. 
In view of this our group has been engaged in investigating 
this particular regime with the help of particle - in - cell (PIC) simulations.  Recently, we illustrated  a possible new mechanism of direct  ion heating by laser pulse \cite{ayushi_EXB}.  For relativistically intense lasers we had also demonstrated the formation of  magnetosonic solitons \cite{kumar_soliton}. In this paper we  show  the direct coupling of the laser energy  to  electrostatic  lower hybrid fluctuations in  plasma.  In addition the   magnetosonic perturbations are also observed in simulation. A detailed parametric study for the frequency regime of the  excitations  of these modes has been carried out.  This paper has been organized as follows. In Sec.II, we have provided the  details of the simulation configuration.   Sec.III contains our numerical observations and analysis.  Finally, Sec.IV provides the summary of the  work.

\section{ Simulation Details}
\begin{figure*}

	\fbox{\includegraphics[width=0.59\linewidth]{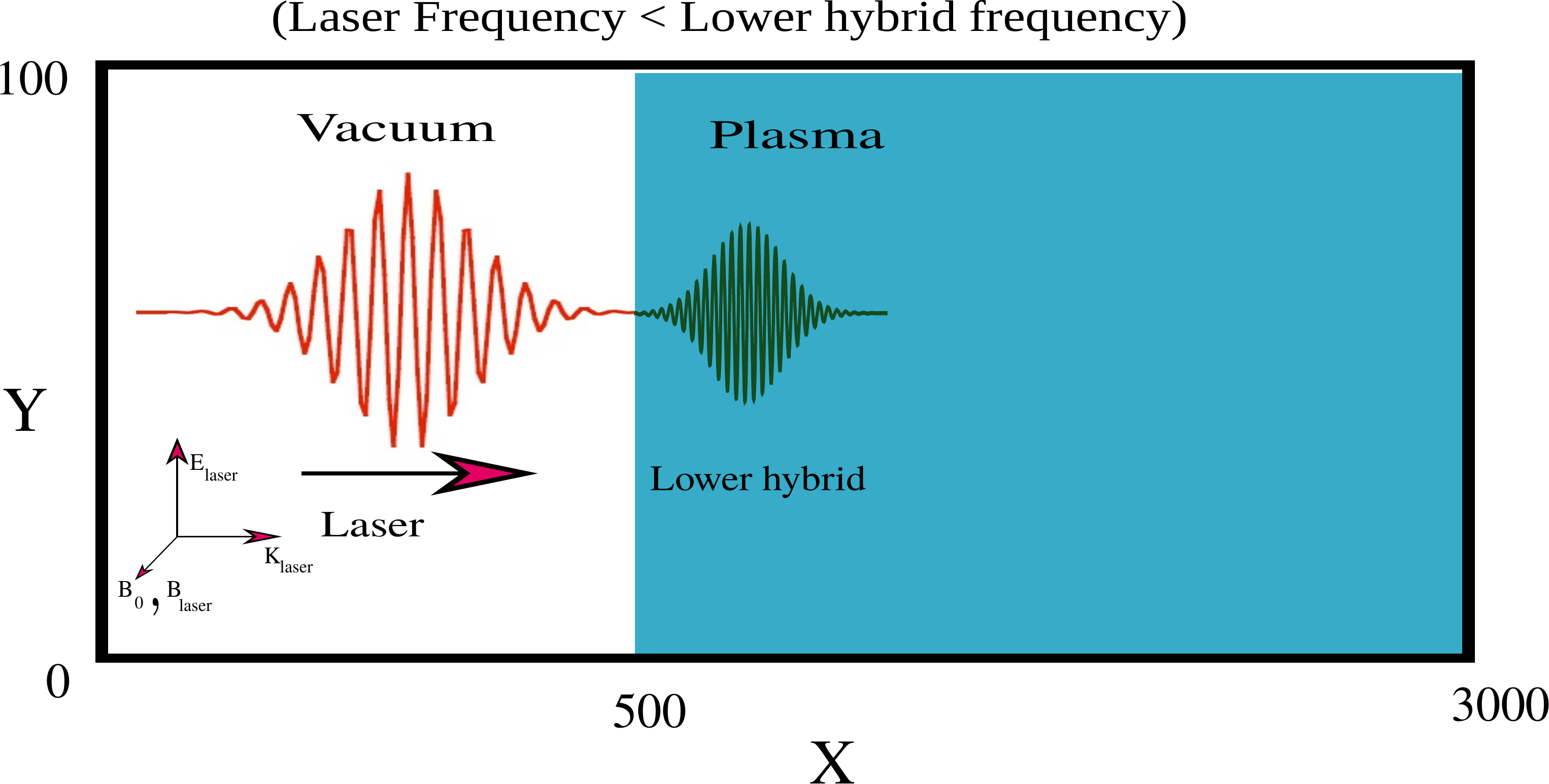}	}	
	\caption{ Schematic showing laser energy being coupled into plasma via excitation of lower hybrid oscillations in the system. Our simulation geometry is in X-Y plane with a planar laser being incident on plasma along $+\hat{x}$. An external magnetic field has been applied in X-mode configuration for the incoming laser. The external magentic field is high enough to magnetise the electrons, keeping ions unmagnetised. For the sake of simplicity, the transverse extent of laser has been kept infinite. }
	
	\label{schematic}
\end{figure*}

We have carried out two dimensional PIC simulations using OSIRIS-4.0 \cite{hemker,Fonseca2002,osiris}. A schematic of the simulation geometry has been shown in Fig.\ref{schematic}.  A rectangular box in x-y plane of dimensions $ L_x= 3000 c/\omega_{pe} $ and  $ L_y = 100 c/\omega_{pe} $ has been chosen for simulation. Here $\omega_{pe} = \sqrt{4\pi n_0 e^2/m_e}$ is the plasma frequency corresponding to the plasma density $n_0$.   The left side of the box upto $x = 500 c/\omega_{pe}$ is vacuum. Thereafter, 
 a uniform density plasma has been placed. 
   A p-polarized plane laser pulse  is incident normally from the left 
 side. We have chosen  the parameters associated with $ CO_2 $ laser pulse having  a wavelength of 
 $10 \mu m$ for our studies. This choice reduces the  requirement on  the applied 
 external magnetic field   by  typically $10$ times  compared to the conventional lasers 
 with  a  wavelength close to $1 \mu m$. This is because the 
   external magnetic field has
     been  chosen so as to have  the lighter electron species magnetized and the ions 
     to remain un-magnetized at the laser frequency. This translates into the requirement of 
     $ \omega_{ce} > \omega_{l} > \omega_{ci}$ where $\omega_l$ is the laser frequency. To satisfy this condition, we have chosen  external magnetic field such that   $  \omega_{ce} = 2.5 \omega_{pe} = 12.5 \omega_l$. The motion of both  electrons and ions 
     are tracked in the simulation. 
 A reduced ion to electron mass ratio of $25$ has been considered. In some cases the mass ratio of $100$ has also been chosen,  which has been explicitly mentioned while discussing those results. The small mass ratio has been chosen  to expedite  the simulations. 
   The boundary condition along $y$-axis for particles as well as fields has been taken to be  periodic.  
   However, along   x-axis the   absorbing boundary condition has been implemented.

   In Table \ref{simulation_parameters} we list out the main simulation parameters for a 
   quick reference.  We have also varied some parameters like magnetic field and the laser frequency in  a couple of our  simulation runs for exploring the conditions of the excitation of these modes. These specific choices have been  specified 
   explicitly 
   where the results concerning them are discussed.

	\begin{table}
		\caption{ Values of simulation parameters in normalised and corresponding standard units for $m_i$=25}
		\begin{tabular}{|p{1.5cm}||p{2.5cm}||p{2.5cm}|}
			\hline
			\textcolor{red}{Parameter}& \textcolor{red}{Normalised Value}& 	\textcolor{red}{Value in standard unit}\\
			
			\hline
			\hline
			\multicolumn{3}{|c|}{\textcolor{blue}{Plasma Parameters}} \\
			\hline
			$n_o$&1 &$3 \times 10^{20}$ $ cm^{-3}$\\
			\hline
			$\omega_{pe}$&1&$10^{15}$Hz\\
			\hline
			$\omega_{pi}$&0.2 (for M/m = 25) &$0.2 \times 10^{15}$Hz\\
			\hline
			\multicolumn{3}{|c|}{\textcolor{blue}{Laser Parameters}} \\
			\hline
			$\omega_l$&$0.2 \omega_{pe}^{-1}$&$0.2 \times 10^{15}$Hz\\
			\hline
			$\lambda_{l}$& $31.4 c/ \omega_{pe}$ &$9.42\mu m$\\
			\hline
			Intensity&$a_{0} =0.5$&$3.5\times 10^{15} W/cm^2$ \\
			\hline
			
			\multicolumn{3}{|c|}{\textcolor{blue}{External Magnetic Field Parameters}} \\
			\hline
			$B_{z}$&2.5 &14.14KT\\
			\hline

			\hline	
				\end{tabular}
					\label{simulation_parameters}
	\end{table}	

	\section{Observations and analysis}
\begin{figure*}
	\includegraphics[width=1.0\linewidth]{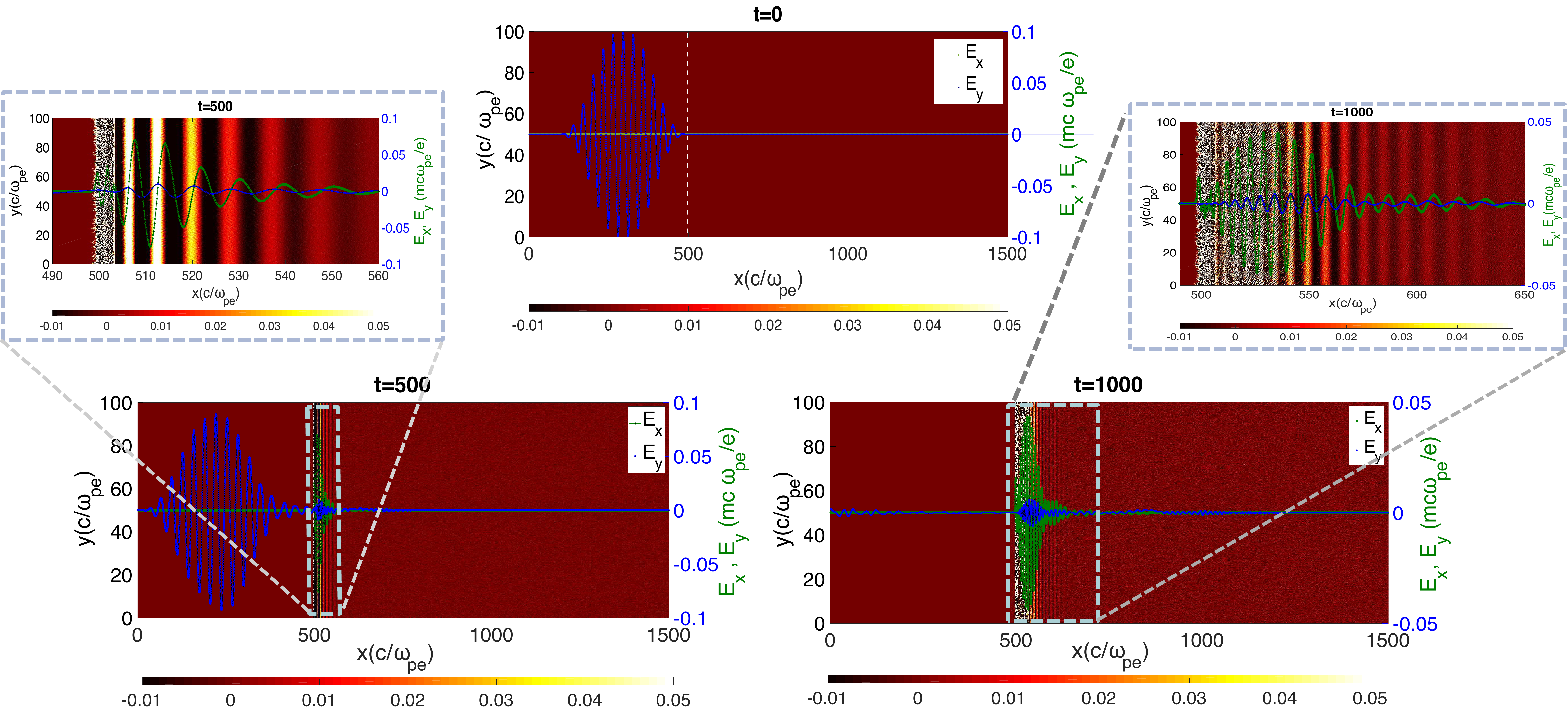}	
	
	\caption{Plot of electric field components ($E_x$ and $E_y$) superposed over 2-D color plot of charge density. At $t=0$, only $E_y$ is present in the system which is due to laser. As laser interacts with plasma ($t=500$), $E_x$ is generated in bulk plasma and spatial variation of $E_x$ is found to be in accordance with that of charge density (zoomed plot of $t=500$ and $t=1000$). This suggests that density perturbations along $\hat{x}$ have generated $E_x$ in bulk plasma.  }
	
	\label{n_colorplot}
\end{figure*}

\begin{figure*}

		\includegraphics[width=0.68\linewidth]{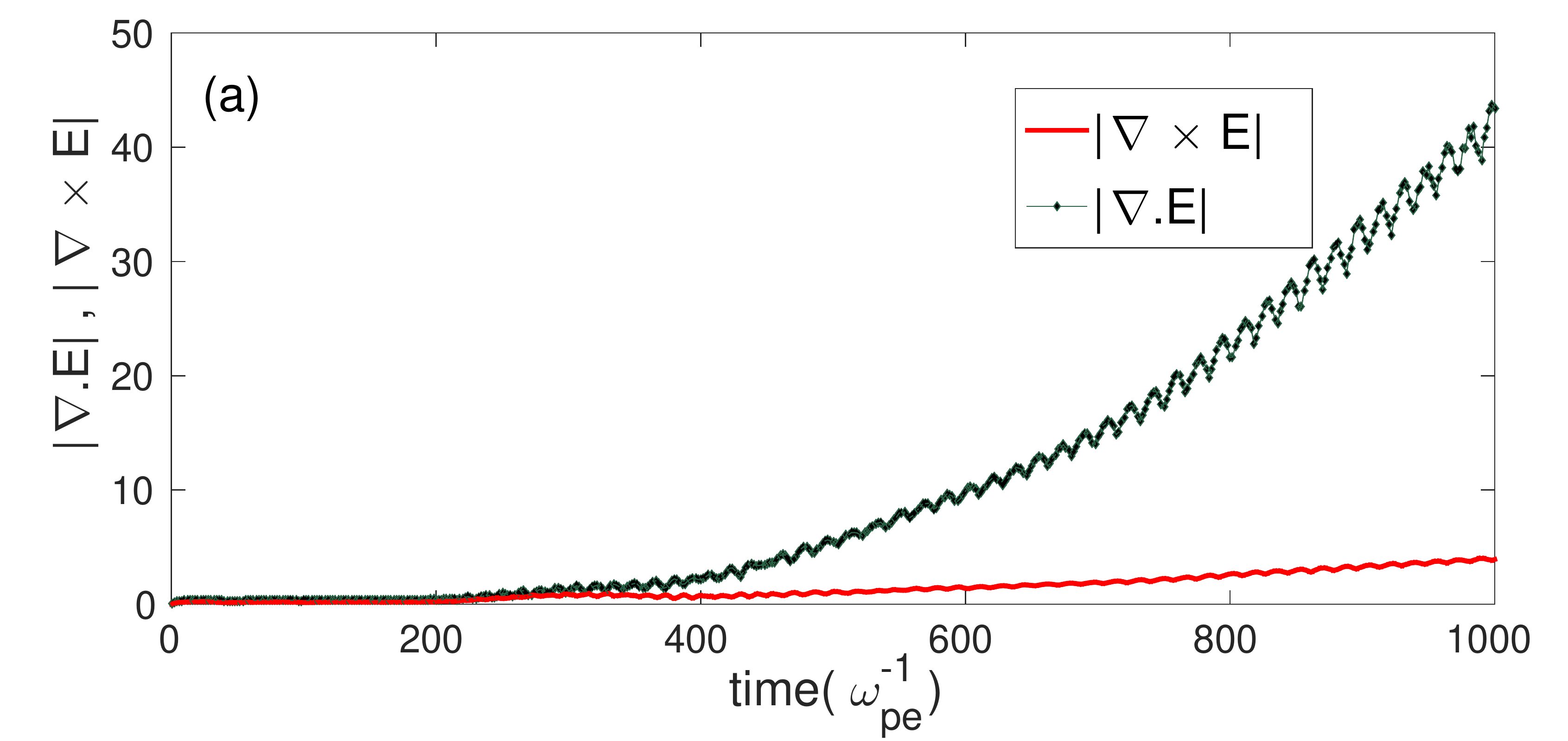}
		\fbox{\includegraphics[width=0.28\linewidth]{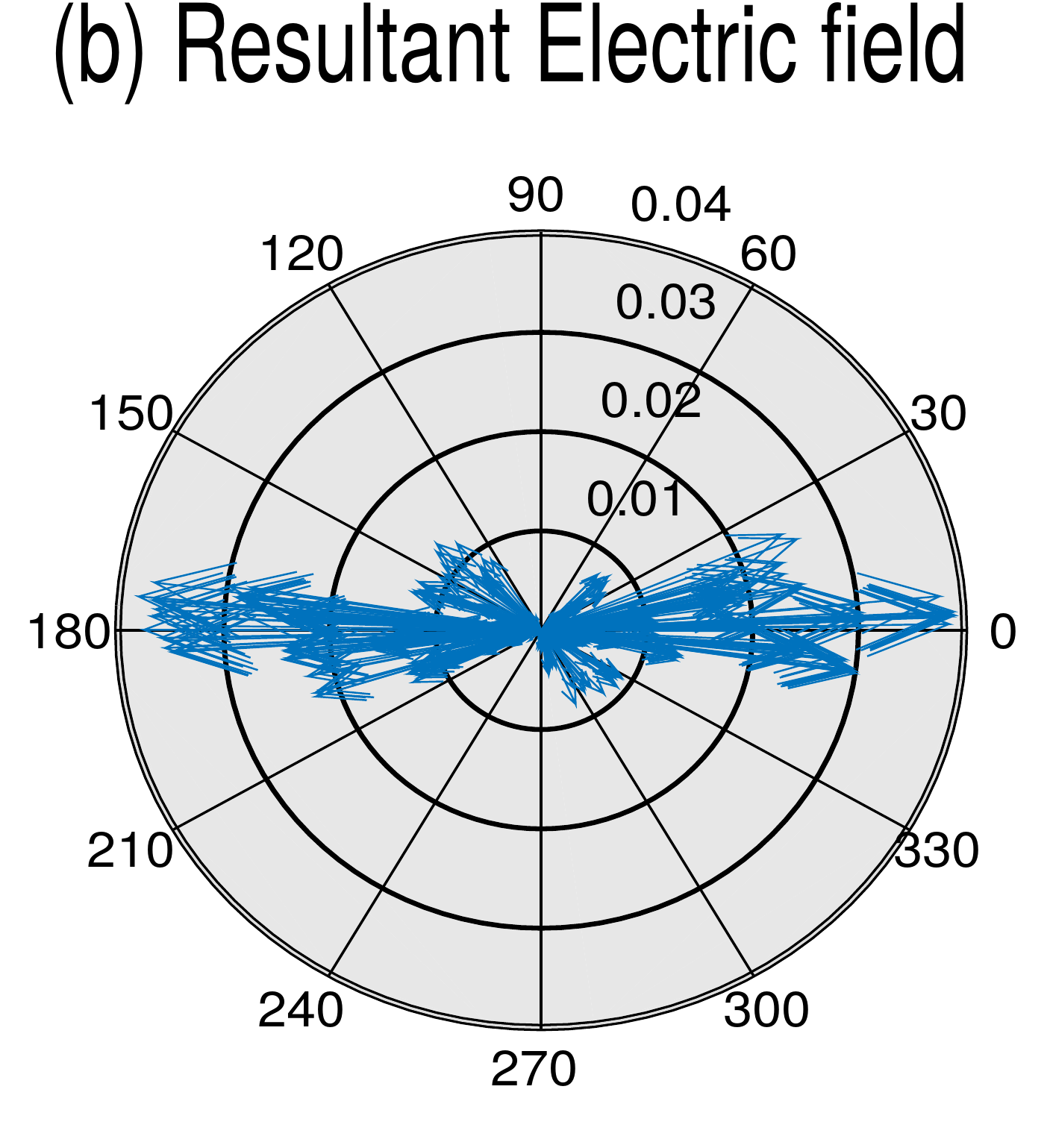}}	
	\caption{(a) Temporal Variation of electrostatic and electromagnetic component of electric field in bulk plasma exhibiting that electrostatic part dominates in bulk plasma. (b) Direction of resultant electric field in bulk plasma at $t=1000$ which is pointing dominantly along $\hat{x}$.}
	
	\label{div_curl}
\end{figure*}

\begin{figure*}
	\includegraphics[width=0.6\linewidth]{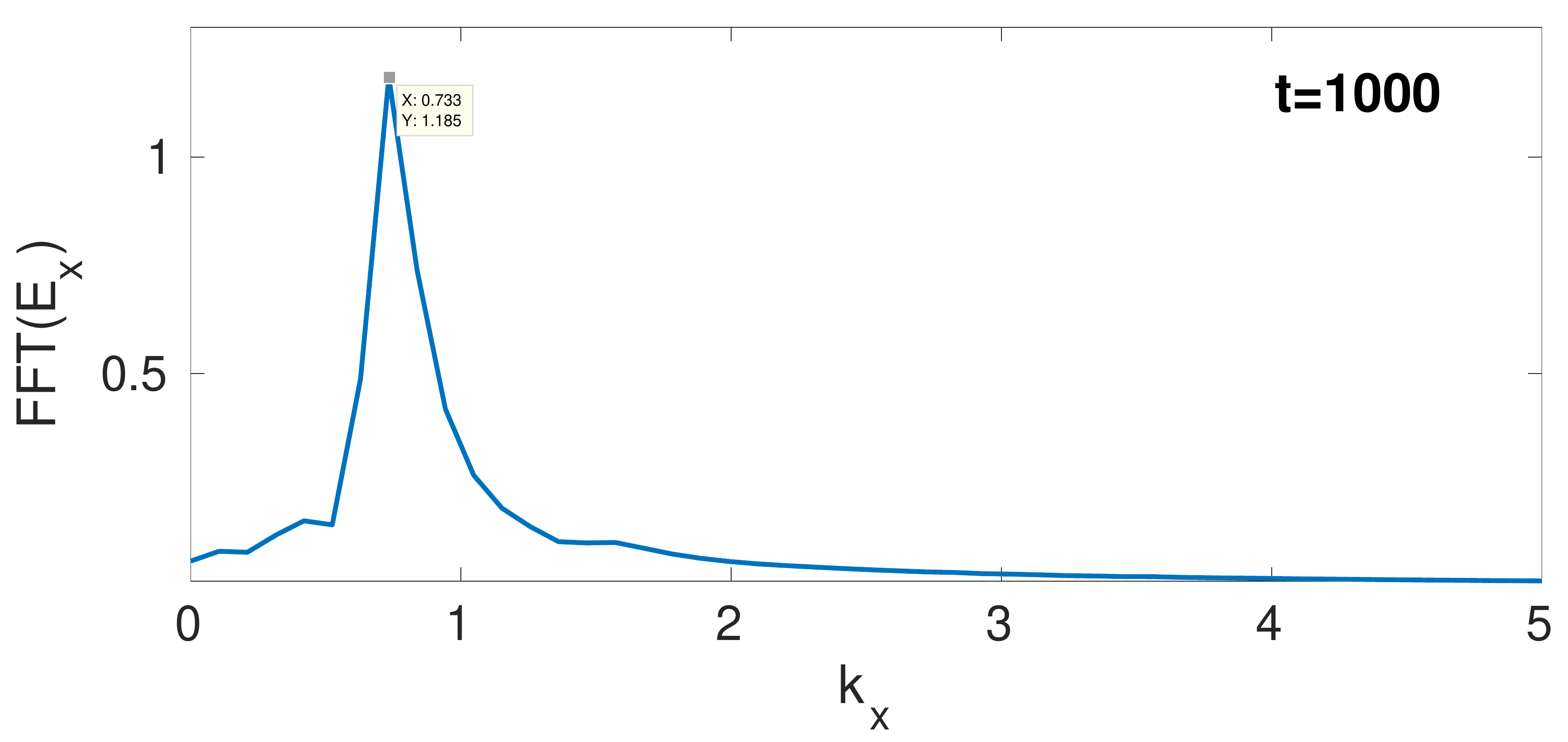}
	
	\caption{ FFT of $E_x$ in bulk plasma along $\hat{x}$ after laser has been reflected back from the system.}
	\label{2d_fft}	
	
\end{figure*}

	From the laser and plasma parameters of Table-\ref{simulation_parameters},  the laser frequency is 
	$\omega_l = 0.2 \omega_{pe}$.  The plasma is overdense for this laser frequency. The laser 
	propagates along $\hat{x}$ axis and is incident normally on the plasma target. It is linearly  polarised with electric field of the laser pointing along $\hat{y}$. 
	 As expected, in the absence of any applied external magnetic field,  the simulation 
	 shows that the laser pulse gets reflected from the plasma target and no disturbance  gets  excited in the plasma medium.  We then carried out simulations in the presence of an external magnetic 
	 field pointing along $\hat{z}$ direction. 
	  The magnetic field was chosen to satisfy the condition $\omega_{ce} > \omega_l > \omega_{ci}$. This ensures  that at the laser oscillation time period, the electrons would exhibit a magnetized response and the ions on the contrary will remain 
	 unmagnetized. The results of these simulations have been presented 
	 in this section.  We observe that with the addition of external magnetic field, 
	 a part of laser energy  gets absorbed by the plasma medium, despite 
	 it being overdense.  This is clearly evident from 
	 Fig. \ref{n_colorplot} where the excited disturbances in the plasma medium have been shown at various times. The plasma surface starts from $x = 500$. The color plot  shows the charge density  in 2-D. 
	 The blue and green lines denote the  $y$ and $x$ component of electric field ($E_y$ and $E_x$) respectively.  At $t=0$ 
	 the plasma is undisturbed and   no charge density fluctuations   can be seen in the medium. The $x$ component of the electric field ($E_x$) is also zero initially. One observes only the $y$ component of electric field ($E_y$) associated with the laser pulse in the vacuum region at $t = 0$. 
	 As the  laser hits the plasma surface, the plasma medium gets disturbed by it. The  effect can be seen  
	  in the subplots of Fig.\ref{n_colorplot} where the plots are shown at   times $t = 500 $ and $t = 1000$.  
	The   charge density disturbances are evident in the  zoomed  plots at these times. 
	 Presence of both components of electric fields  $E_x$ and $E_y$   in the plasma medium can be seen.  
	  With time, these disturbances  propagate inwards,  towards  the deeper region of the 
	  plasma medium (comparison of plots at $t=500$ and $t = 1000$ illustrates this). 
      	  These disturbances are not random fluctuations but appear to be quite regular.
     In Fig. \ref{div_curl}(a),  we have plotted the evolution of $I_1 =\int \mid \nabla \cdot \vec{E} \mid  dx dy$ and 
     $I_2 = \int \mid \nabla \times \vec{E} \mid dx dy$ (integrated over the  bulk region of the plasma) to 
     determine the electrostatic/electromagnetic character of these disturbances.  It can be observed that though both $I_1$  and $I_2$ start with zero initially, they
     increase with time.  However, $I_1$ increases much more rapidly and clearly dominates over $I_2$. 
     This suggests that the fluctuations have a dominating  electrostatic 
     character. This is further borne out from  Fig. \ref{div_curl}(b) where we show directional distribution of $\vec{E}$ in the plasma region.  It demonstrates that the 
     electric field   $\vec{E}$ is dominantly along $\hat{x}$. We have also  evaluated the spatial FFT of 
     $E_x$ in bulk plasma region at $t=1000$ (when the laser has been reflected back from the system) and observe that the spectrum peaks at a particular value of $k_x = 0.73$ (fig. \ref{2d_fft}). 
        The  scale length of the fluctuations appear to be longer  in the deeper region of the  target as can be observed in Fig.\ref{n_colorplot}. 

\begin{figure*}
	\centering
	{\color{magenta}\textbf{\underline{Laser frequency lying in Region I}}}\par\medskip
	\includegraphics[width=0.49\linewidth]{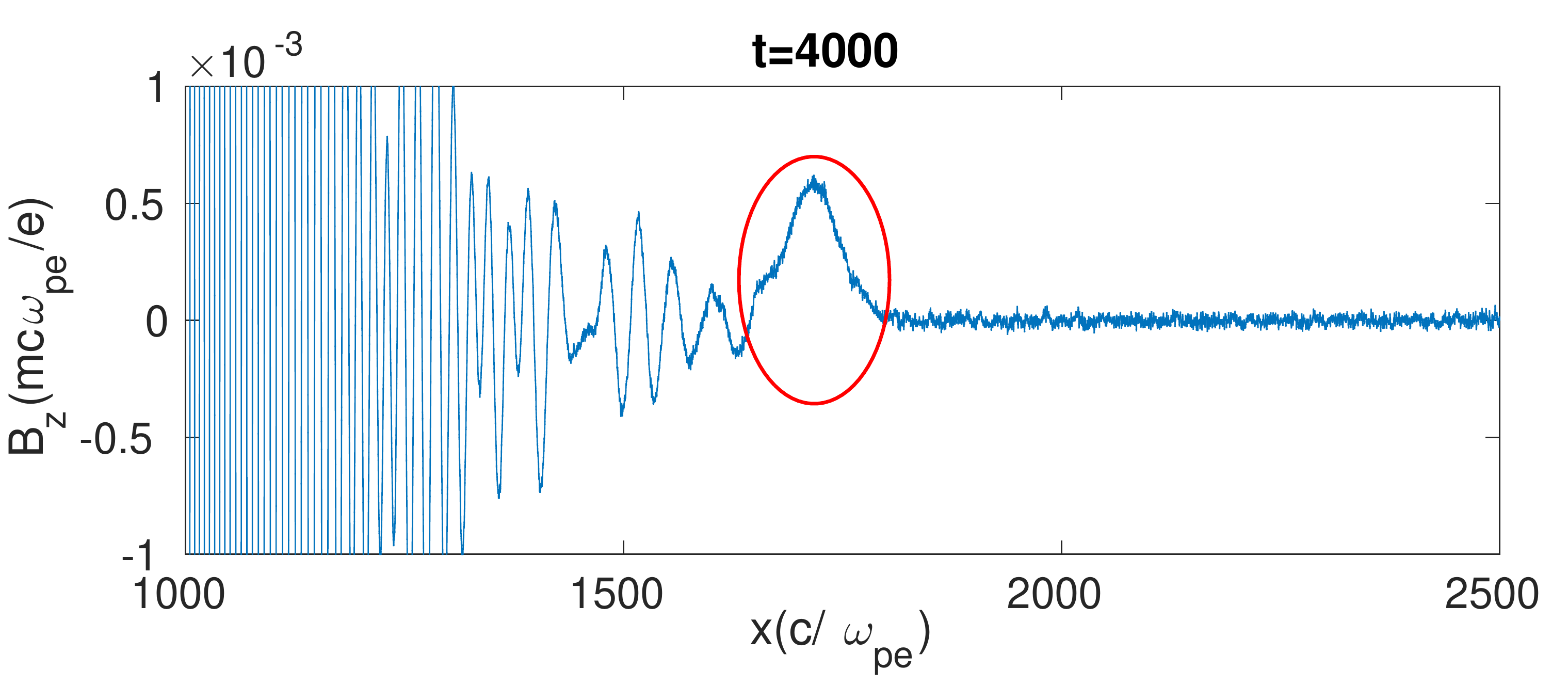}	
	\includegraphics[width=0.49\linewidth]{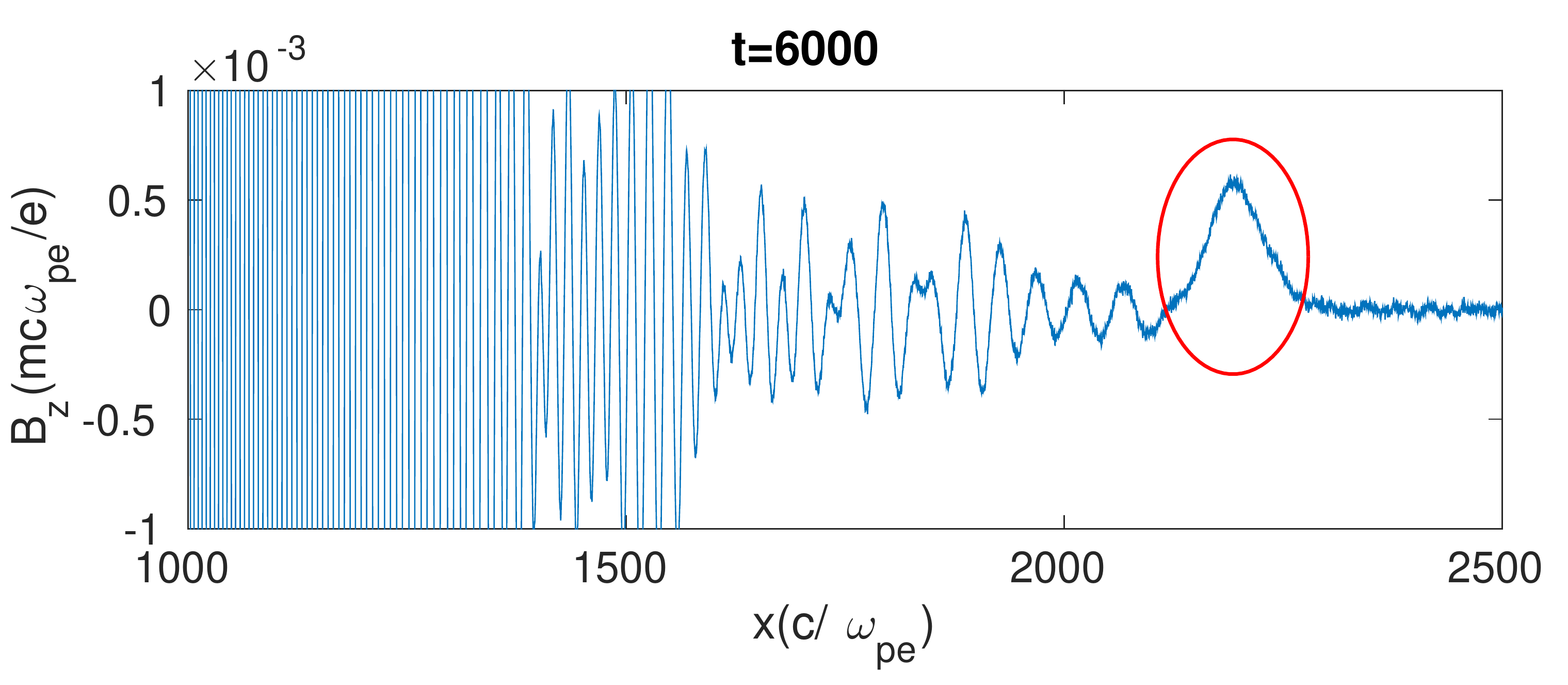}	
	
	\centering
	{\color{magenta}\textbf{\underline{Laser frequency lying in Region II}}}\par\medskip
	\includegraphics[width=0.49\linewidth]{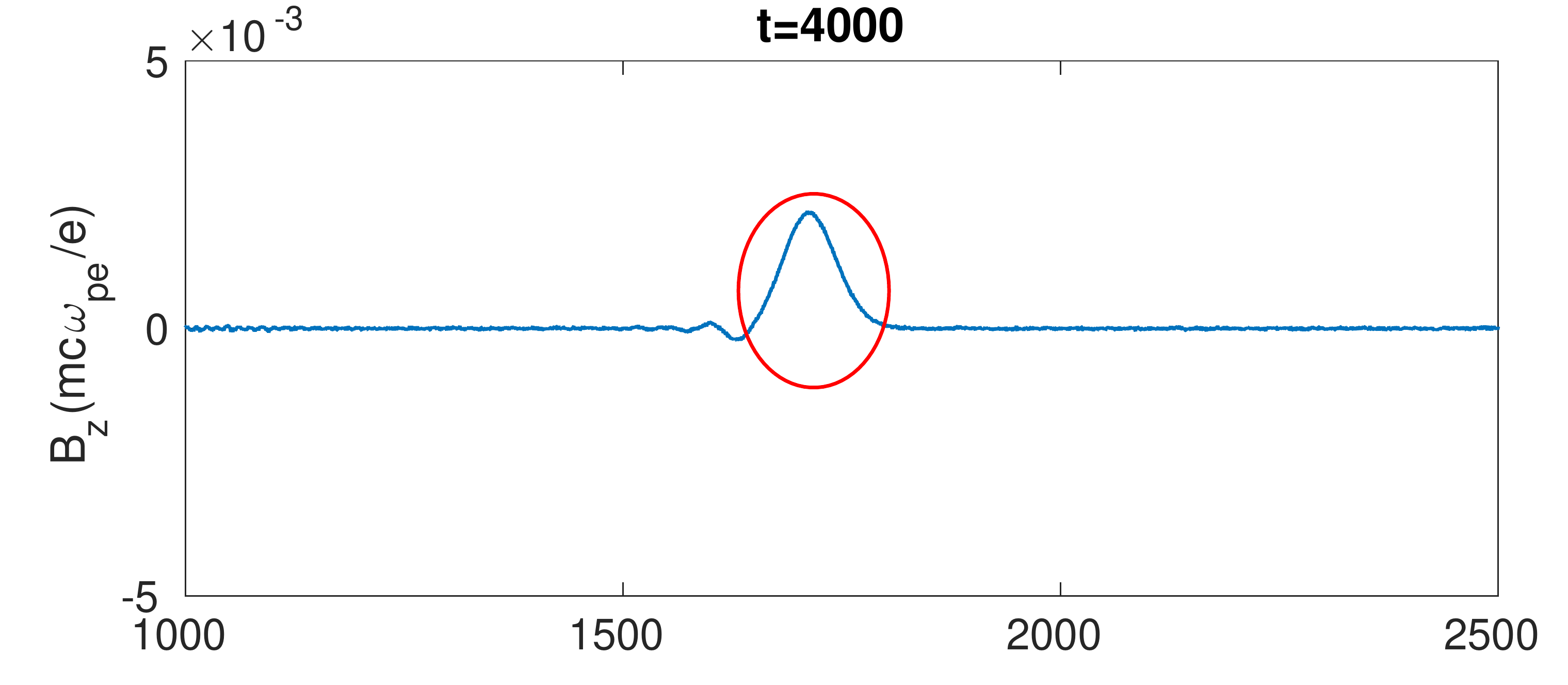}	
	\includegraphics[width=0.49\linewidth]{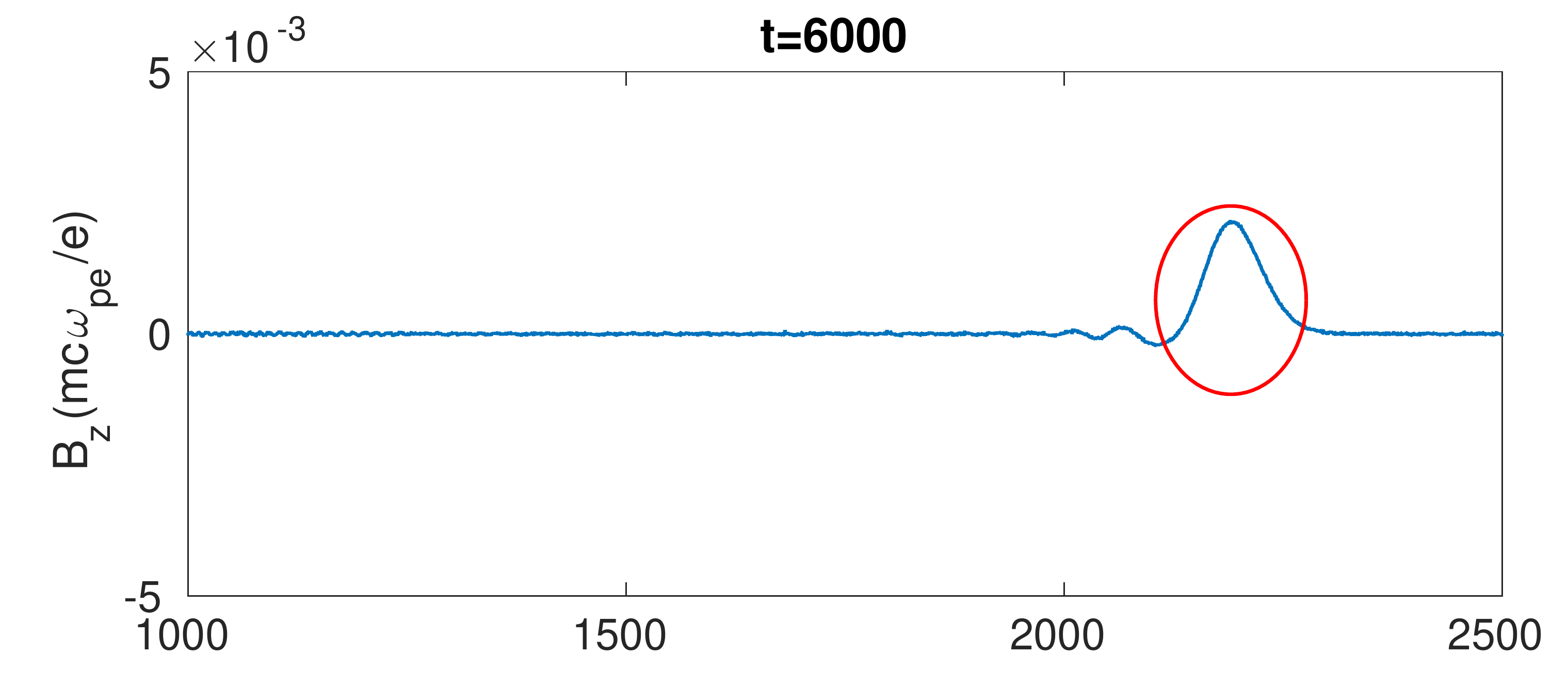}	
	
	\caption{ Spatial variation of $B_z$ at different times. We observe ion heating as well as a structure ahead of the oscillations for laser frequency lying in Region I of frequency band, on the other hand, only a structure for laser frequency lying in Region II.   }
	
	\label{soliton_heating}
\end{figure*}

At a significantly later time (shown at $t = 4000$  and $t = 6000$ in Fig. \ref{soliton_heating}) formation of another structure 
quite distinct from the oscillations discussed so far can be seen clearly in the plot.  This structure is  significantly   longer compared to the oscillations observed at earlier times (Fig. \ref{soliton_heating}). 
  We now make an attempt at understanding the excitation and detail characterization of these two kind of structures in the subsections below.

\subsection{Identification of  short scale fluctuation as  lower  hybrid mode} 
We now show that the short scale electrostatic disturbance generated in the plasma is essentially the 
lower hybrid mode. We also illustrate the physical mechanism and the condition that need to be satisfied for such excitations. 

\begin{figure*}
	\includegraphics[width=0.48\linewidth]{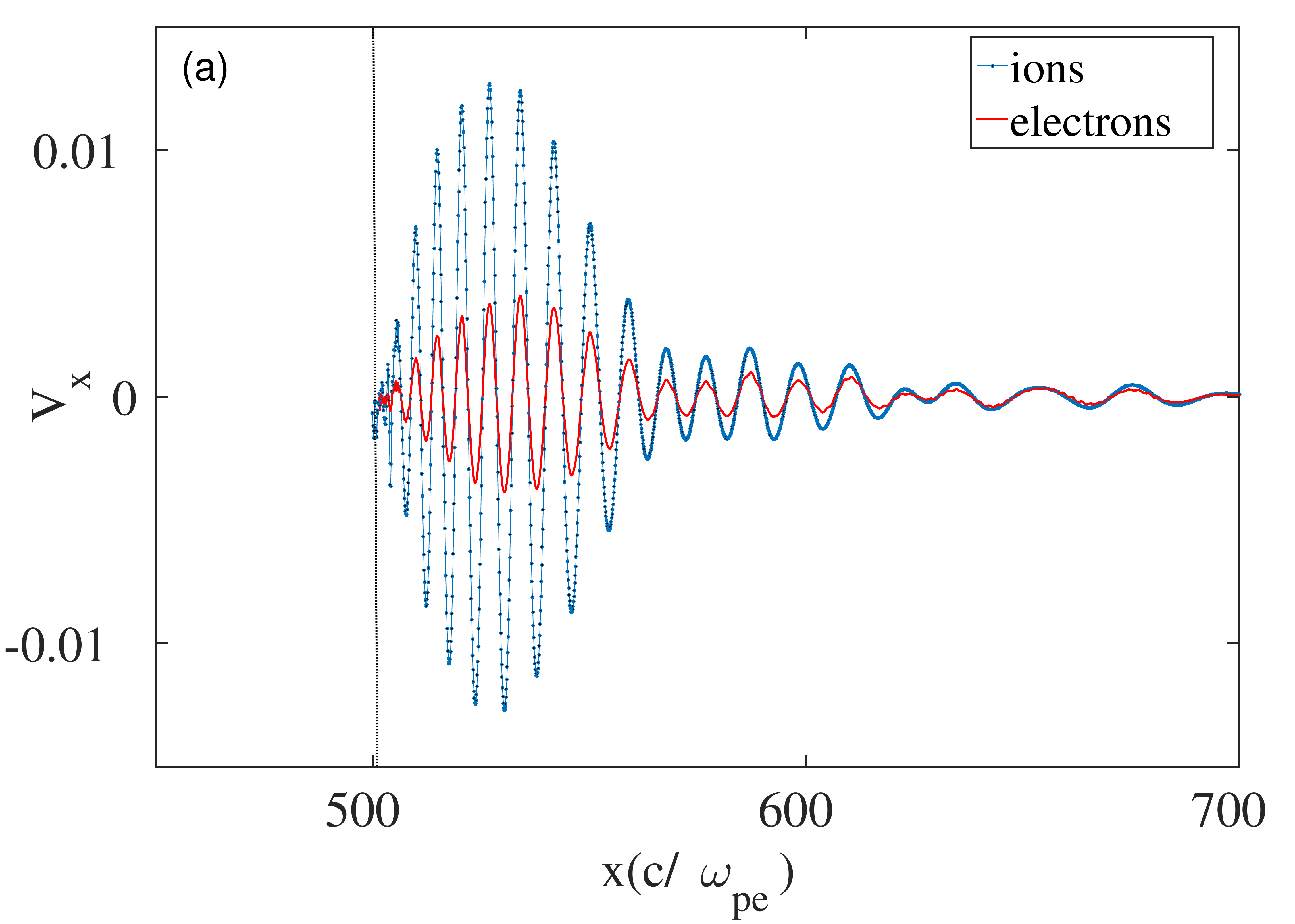}	
		\includegraphics[width=0.49\linewidth]{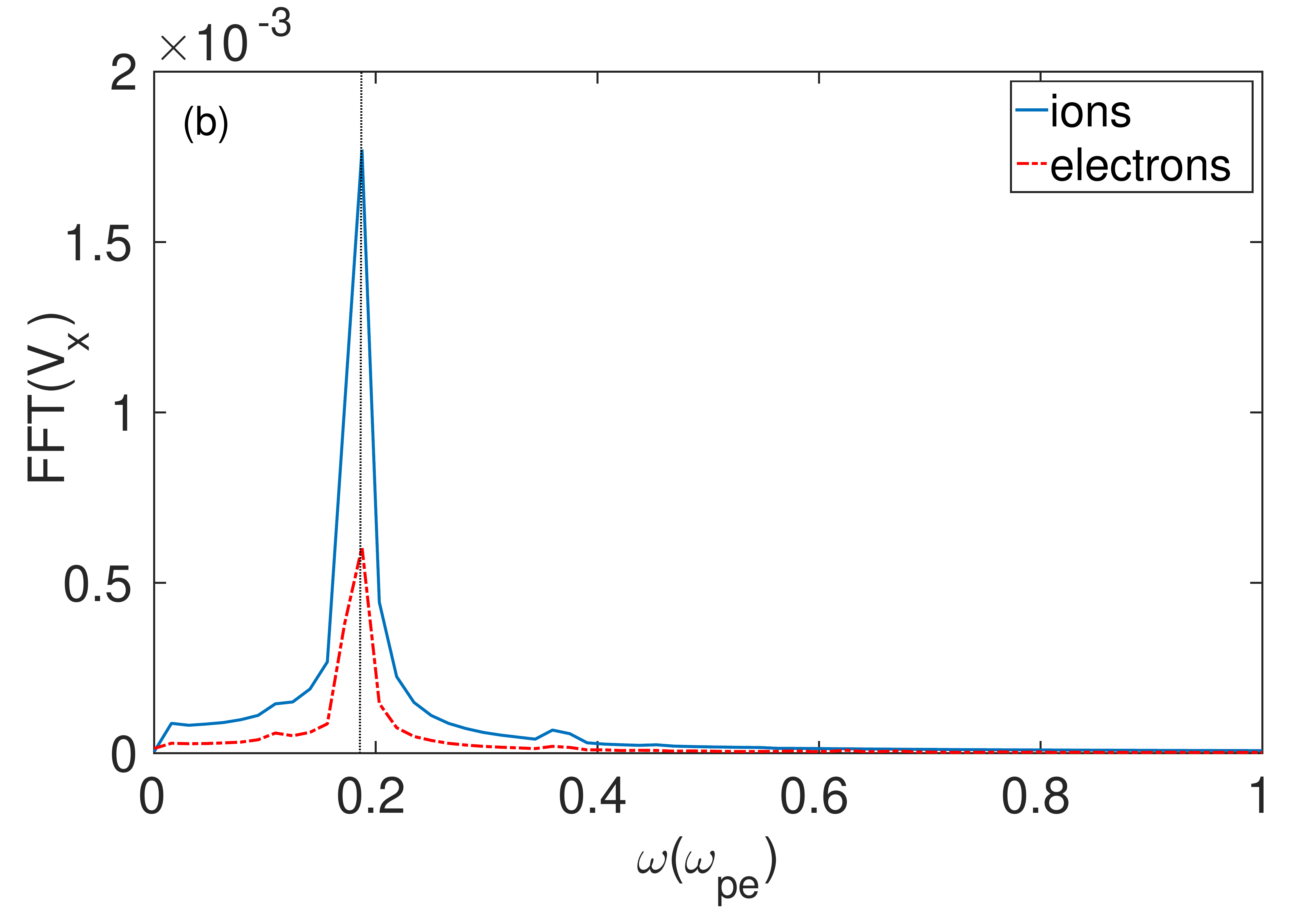}

	\caption{(a) Spatial variation of $V_x$ for both the charged species showing that the magnitude of $V_x$ for ions is more than that for electrons but their motion is in correspondance with each other. (b) FFT of $V_x$ of both the charged species with time again confirms that the frequency of their motion is same \emph{i.e} they move together, indicating  presence of a hybrid mode in the system. }
	
	\label{vx}
\end{figure*}

It should be noted that  in the presence external magnetic field and the oscillating electric field of the laser, both ions and electrons would  experience  $\vec{E} \times \vec{B}$ drift along  $\hat{x}$. 
Here $\vec{E}$ is the electric field  of the laser pulse along $\hat{y}$
and $\vec{B}$ is the applied external magnetic field. Since the electric field of the laser is oscillating in time the magnitude of this   drift  is dependent on the charge species and takes the following form \cite{Nishikawa}
\begin{equation} 
\label{EXB}
\vec{V}_{s,\vec{E} \times  \vec{B}}(t) = \frac{ \omega_{cs}^2}{ \omega_{cs}^2 - \omega_{l}^2} \frac{ \vec{E}(t) \times \vec{B}}{B^2}
\end{equation}
  	Here, the suffix $s = e, i $ stand for  electron and ion species respectively. Thus, $\omega_{ce}$ and $\omega_{ci}$ denote  the electron and ion cyclotron frequency respectively and $\omega_l$ is  the laser frequency.
We have maintained the condition of $ \omega_{ce} > \omega_{l} > \omega_{ci}$ in all our simulation studies here. 
The difference in this drift speed between the two species 

gives rise to a current in the plasma medium, viz.,  
$\vec{J} = e n_0 (\vec{V}_{i,\vec{E} \times  \vec{B}}(t)  - \vec{V}_{e,\vec{E} \times  \vec{B}}(t) ) $. 
 This current is directed along  $\hat{x}$. The spatial variation of laser electric field along  $x$ at laser wavelength makes the 
$\vec{\nabla} \cdot \vec{J} (= \partial J_x/\partial x)$ finite. The  
finite divergence of the current 
 leads  to space charge  fluctuation from the 
continuity equation. This charge density fluctuation is responsible for  the  creation 
of  the electrostatic field $E_x$. Once $E_x$ sets up the two charge species also respond to the 
force ($q E_x$) along the $\hat{x} $. 
 However, the electrons being magnetized, they do not accelerate 
freely by $E_x$.  On the other hand ions being un-magnetized get directly accelerated by this field. 
It is observed in the simulations that dominant $x$ component of the velocity for electron 
is provided by the expression of  $\vec{E} \times \vec{B}$ drift given in Eq.(\ref{EXB})
in  the laser electric field, for the ions on the other had the velocity gained by 
direct acceleration through $E_x$ dominates. The plot in Fig.\ref{vx} shows the spatial variation of the 
$x$ component of velocity of the two species obtained from simulations. 
The two species move together, however, the  amplitude of their velocity differs, with the ions showing 
higher amplitude. 

The frequency spectrum of the observed oscillation has been shown in the subplot alongside. The spectrum peaks 
at the frequency of $\omega = 0.176$ which differs from the laser frequency of $0.2$.    It matches closely with the 
  lower hybrid frequency  of $0.186$ evaluated from the analytical expression of
 \begin{equation}
\label{lh_dispersion}
\omega_{LH} = {\left({\frac{1}{\omega_{pi}^{2}} + \frac{1}{\omega_{ce}\omega_{ci}}} \right)}^{-1/2} 
\end{equation} 
  It is thus clear that the laser excites an electrostatic mode in the medium in which both electrons and ions have significant roles to play.

\subsection{Necessary condition for the generation of LH mode}
We have shown  that the generation of the  electrostatic field is linked to the difference in the $\vec{E} \times \vec{B}$ field of the two species in the presence of external 
magnetic field and  the oscillating transverse electric field of the laser. Clearly, this requires 
that the laser field should penetrate the  plasma. It should be noted that we have chosen 
the plasma medium to be  overdense for the chosen laser frequency   in  our simulations here. 
Thus the laser field in the normal course would not have penetrated the plasma region.

\begin{figure*}
	
	\fbox{\includegraphics[width=0.6\linewidth]{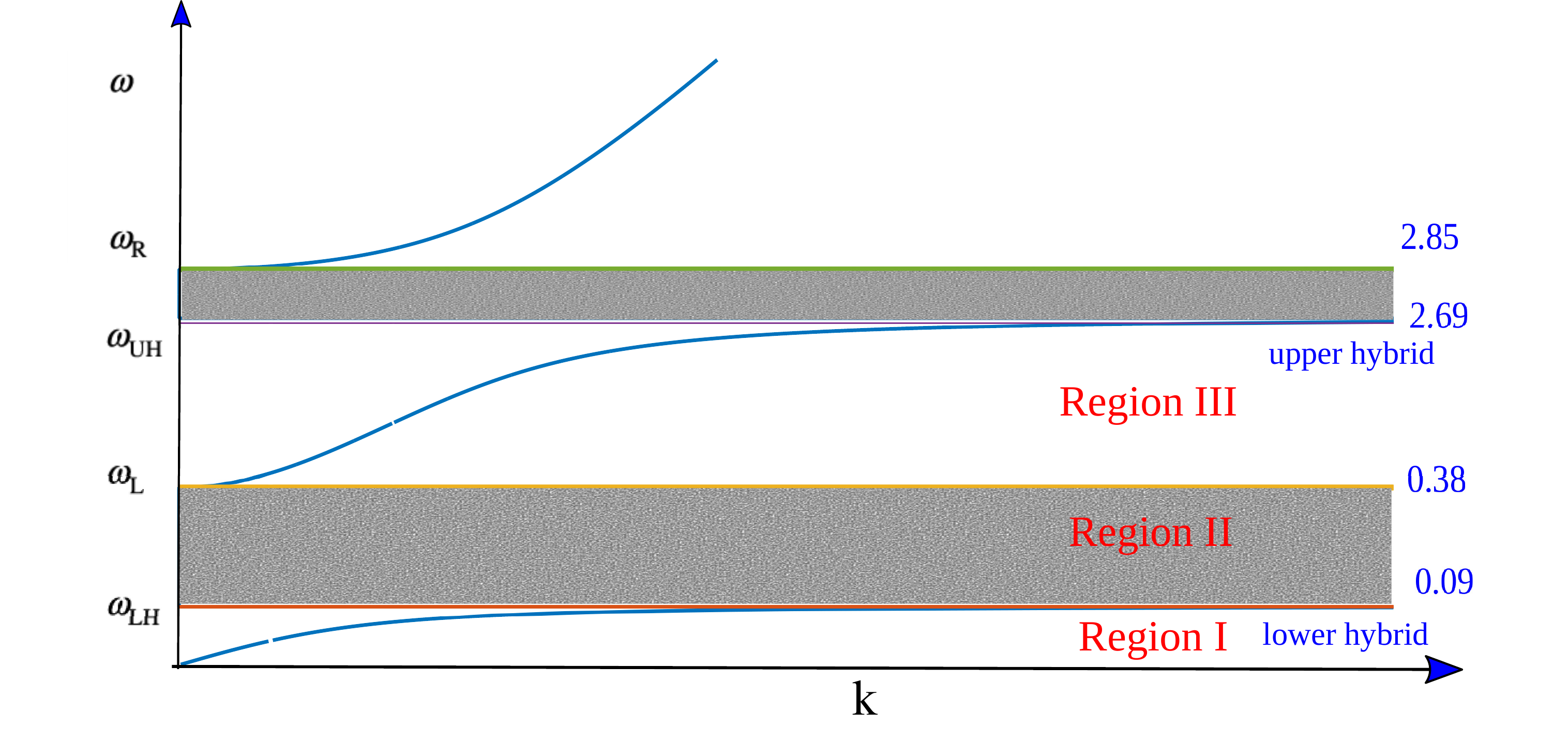}}	
	
	\caption{ Dispersion curve for X-mode indicating different stop band and pass band frequencies (Fig. courtesy \cite{boyd_sanderson_book}). Value of different frequencies (normalised to $\omega_{pe}$) corresponding to $m_i =100$ and $B_z=2.5$ is shown at right in blue.   }
	
	\label{dispersion}
\end{figure*}

However, the laser radiation is able to penetrate the plasma despite having frequency lower than the electron plasma frequency due to the presence of  applied external magnetic field. 
The oscillatory electric field of the electromagnetic field is  directed 
orthogonal to the applied magnetic field. This  suggests 
 that the  $X$ mode is the relevant  mode for our simulation geometry. 
The dispersion curve of the $X$ mode has the 
 characteristics shown in Fig.\ref{dispersion} (adapted from Boyd and Sanderson \cite{boyd_sanderson_book}).  
 The figure  shows 
  the presence of  a stop band  between $\omega_{LH}$ and $\omega_L$ indicated by 
  the shaded region, which has been  denoted as Region II in the figure. 
  Region I in frequency band ranges from $0$ to $\omega_{LH}$ and is the pass band, so is 
  Region III from $\omega_L$ to $\omega_{UH}$ for the incoming electromagnetic wave. 
  Here, $\omega_{LH}$ is the lower hybrid frequency and has been defined in equation \ref{lh_dispersion} , $\omega_{UH}=\sqrt{\omega_{pe}^2 + \omega_{ce}^2} $ is the upper hybrid oscillation frequency.

The frequencies $\omega_L$ and $\omega_R$ shown in Fig. \ref{dispersion} corresponds to the left and right hand cut off defined by the following expressions
	\begin{equation}
\label{wl}
\omega_L= [\omega_{pe}^2+\omega_{pi}^2+(\omega_{ci}+\omega_{ce})^2/4]^{1/2} -(\omega_{ci}-\omega_{ce})/2
\end{equation}	
\begin{equation}
\label{wr}
\omega_R= [\omega_{pe}^2+\omega_{pi}^2+(\omega_{ci}+\omega_{ce})^2/4]^{1/2} +(\omega_{ci}-\omega_{ce})/2
\end{equation}	
  	
On the basis of this distinction between the three regions we now identify the necessary  condition for the excitation of the lower hybrid mode. We choose three different 
values of the laser frequency corresponding to these three regions. To have a better distinction between 
the three regions for these simulation runs, we have chosen the ratio of  ion to electron mass to be  $100$.  This helps in having the three frequencies 
 $\omega_{l}$, $\omega_{ce}$ and $\omega_{ci}$ to be well separated. 
 Since, the physics that needs to be addressed can be explored in a simple 1-D geometry 
 we have chosen the same for these set of runs. We have already,  in all our earlier studies,  
 shown that  2-D effects  do not alter  any physics associated with the main theme of the investigation 
 \cite{ayushi_EXB}.  The simulation box in 1-D geometry is chosen to be larger with $L_x=4000c/\omega_{pe}$ and the simulation duration was also increased. We chose to give simulation runs for three different values of laser frequency corresponding to the  three  different regions  in the $X$ mode dispersion curve (marked as region I, II and III in fig. \ref{dispersion}). The three values of frequencies are $\omega_{l1}=0.08 \omega_{pe}$, $\omega_{l2}=0.16 \omega_{pe}$ and $\omega_{l3}=0.5 \omega_{pe}$ falling in region I, II and III respectively.  For our given simulation parameters $\omega_L$, $\omega_{LH}$ and $\omega_{pi}$ are $0.376$, $0.093$ and $0.1$ respectively. The  stop-band for propagation of the $X$ mode  electromagnetic wave in plasma lies between $\omega_L$ and $\omega_{LH}$. Thus we expect that while the laser would propagate inside plasma for runs with frequency $\omega_{l1}$(pass band) and $\omega_{l3}$(pass band), there should be total reflection of the laser for frequency lying in $\omega_{l2}$(stop band). This is indeed 
   what we observe in the simulations. Since there is no propagation of the laser field at $\omega_{l2}$ 
 the laser does not interact with plasma and  it is not possible to excite lower hybrid oscillations in this particular case. On the other hand, the other two frequencies,  
    $\omega_{l1}$ and $\omega_{l3}$ lie in the pass-band and hence are expected to  interact with 
    the plasma. 

\begin{figure*}
		
		\includegraphics[width=0.5\linewidth]{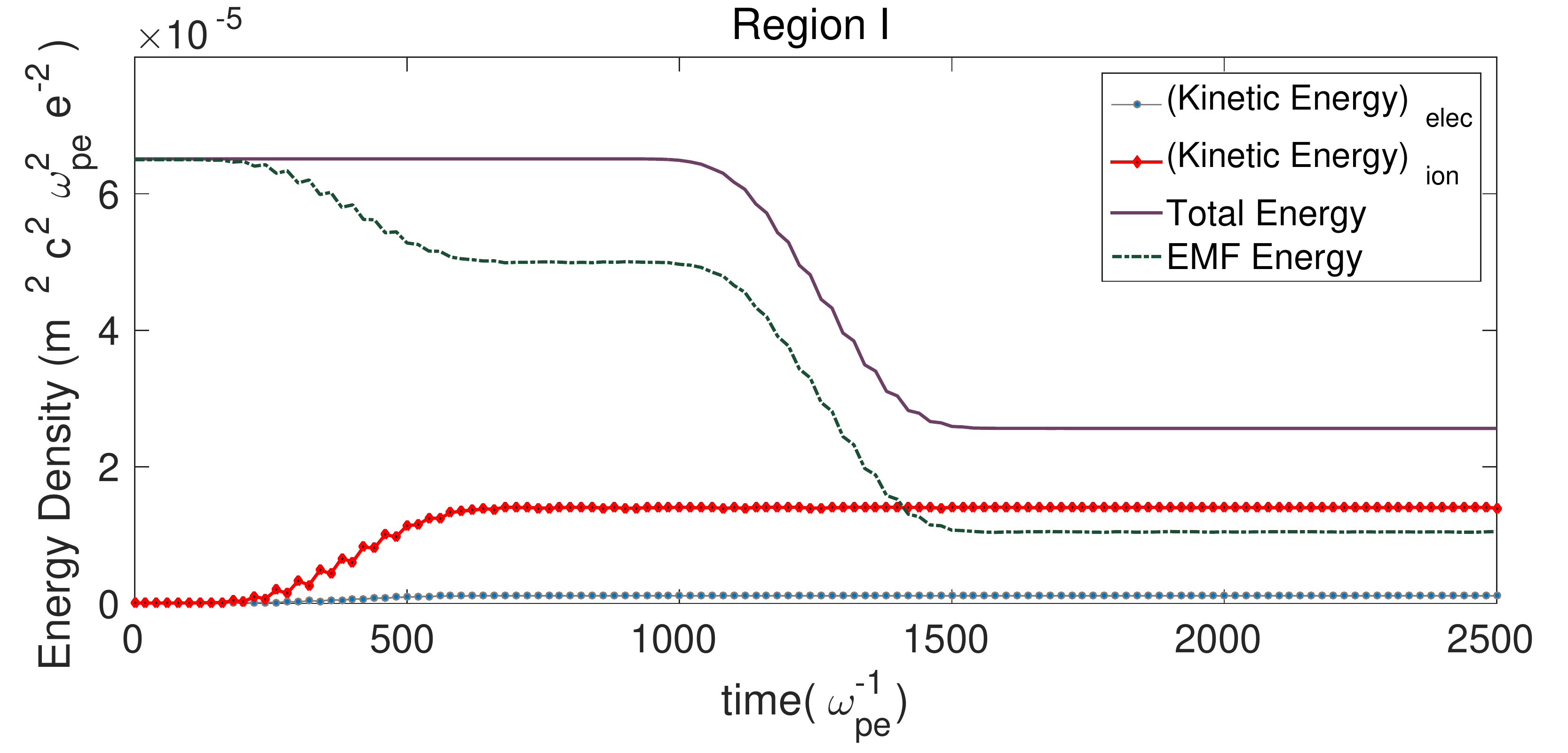}	
		\includegraphics[width=0.5\linewidth]{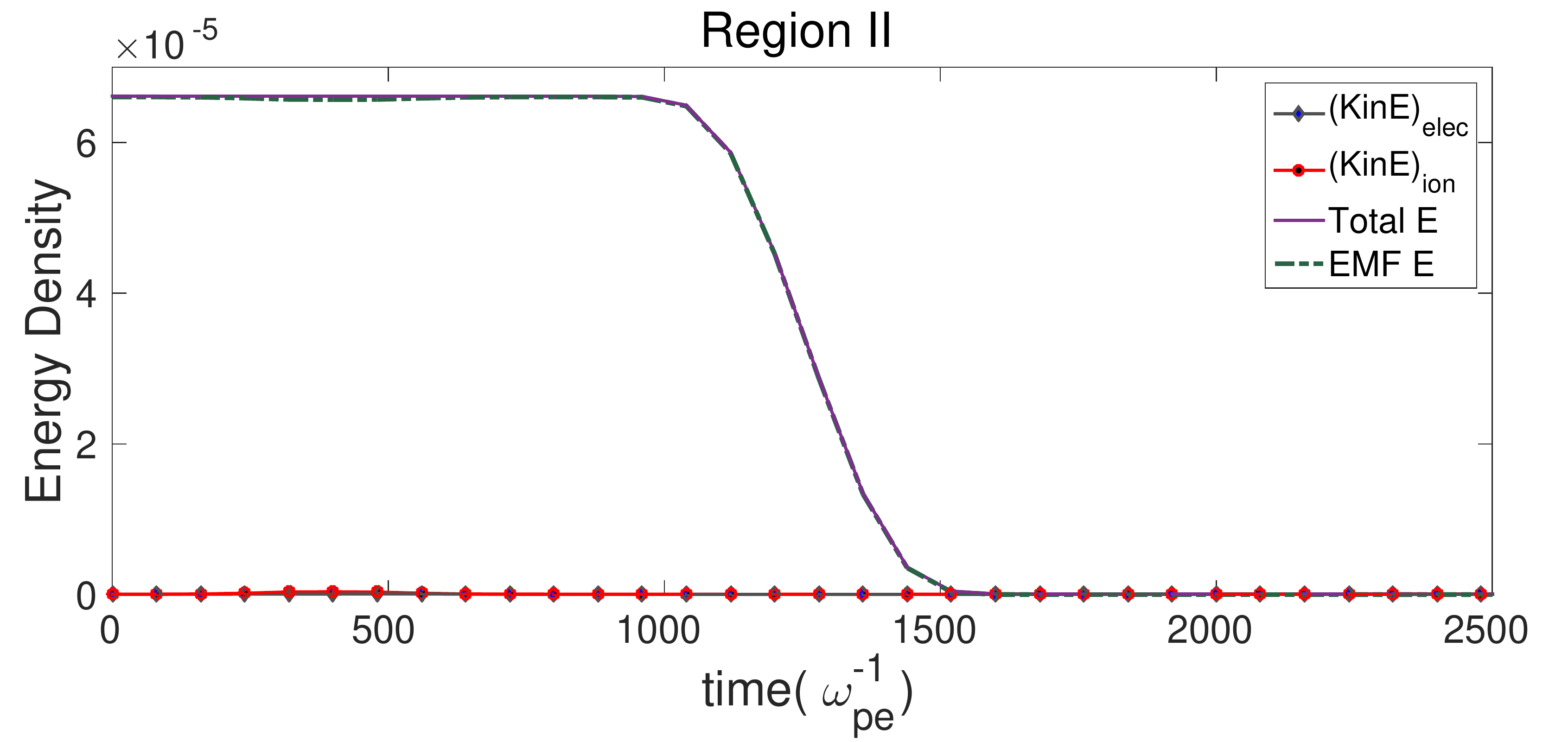}	
		\includegraphics[width=0.5\linewidth]{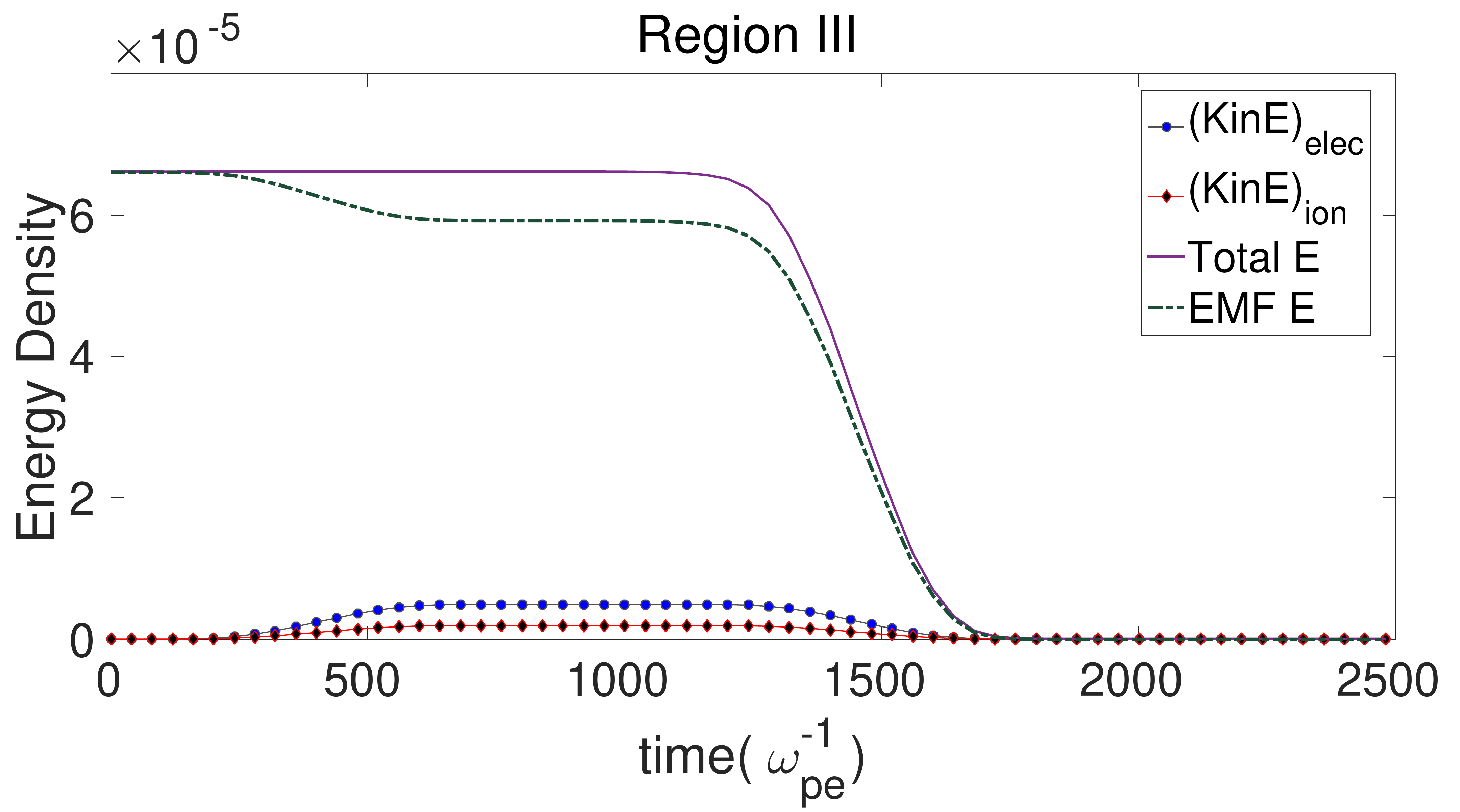}	
	\caption{ Absorption of laser energy into plasma when laser frequency falls into different regions of frequency spectrum. It can be observed that laser energy is coupled into plasma only in Region I of laser frequency. Laser frequency in Region II falls in stop-band and hence, is not able to interact with plasma. On the other hand, laser frequency in Region III lies in pass-band and hence could interact with plasma but could not impart its energy into plasma species. Plasma species acquire energy only till the time laser is present in this case, their energy goes off after that. }
	
	\label{energy}
\end{figure*}

 We, however,  observe that the  laser energy gets  coupled into plasma only when the  laser frequency 
 lies in region I, where the electrostatic lower hybrid wave get excited by the mechanism discussed above. 
 This is because the lower hybrid frequency is nearby and therefore gets  excited. In region III 
 where the laser frequency is much higher than the lower hybrid frequency the coupling of laser energy is found to be very weak with the plasma. The laser simply gets transmitted in this particular case. 
 This can be clearly seen from the evolution of 
 kinetic and field energies shown in Fig. \ref{energy} for frequencies lying in the three regions.

 It thus appears that to excite the lower hybrid wave we need to choose the frequency of the 
 driving laser pulse to be in region I, i.e. lower than the lower hybrid resonance to satisfy the pass 
 band criteria. Furthermore, it should also not be in Region III which satisfies the pass band criteria but the laser frequency is much higher than the lower hybrid wave frequency to couple with it.

\begin{table}
	\caption{Frequencies and their respective normalised values for $m_i=100$ described in for section C of results}.
	\begin{tabular}{|p{1.6cm}||p{1.6cm}||p{2.6cm}|}
		
		\hline
		{Frequency}& {Normalised value} &{Observation}\\
		
		\hline
		\hline
		$\omega_{l1}$ (Region I)&0.08&	absorption via excitation of LH\\
		\hline
		$\omega_{l2}$ (Region II)&0.16&stop-band, hence, no absorption\\
		\hline
		$\omega_{l3}$ (Region III)&0.5&pass band but no coupling of laser energy into plasma\\
		\hline

		\hline

	\end{tabular}
	\label{table_regions}
\end{table}
\subsection{Identification of long scale disturbances as Magnetosonic excitation}
\begin{figure*}
			\centering
	{\color{magenta}\textbf{\underline{Laser frequency lying in Region II}}}\par\medskip
	
		\includegraphics[width=0.59\linewidth]{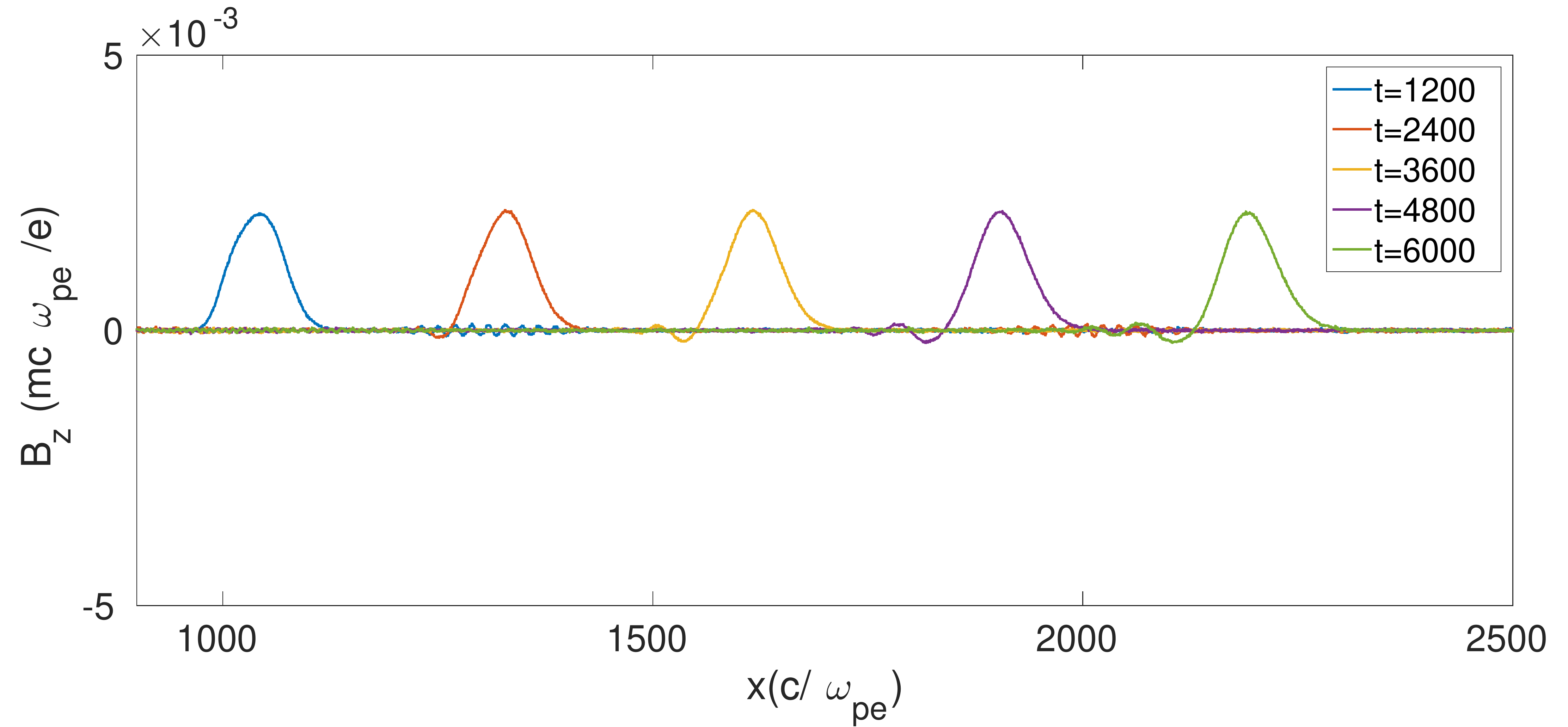}	
		\includegraphics[width=0.59\linewidth]{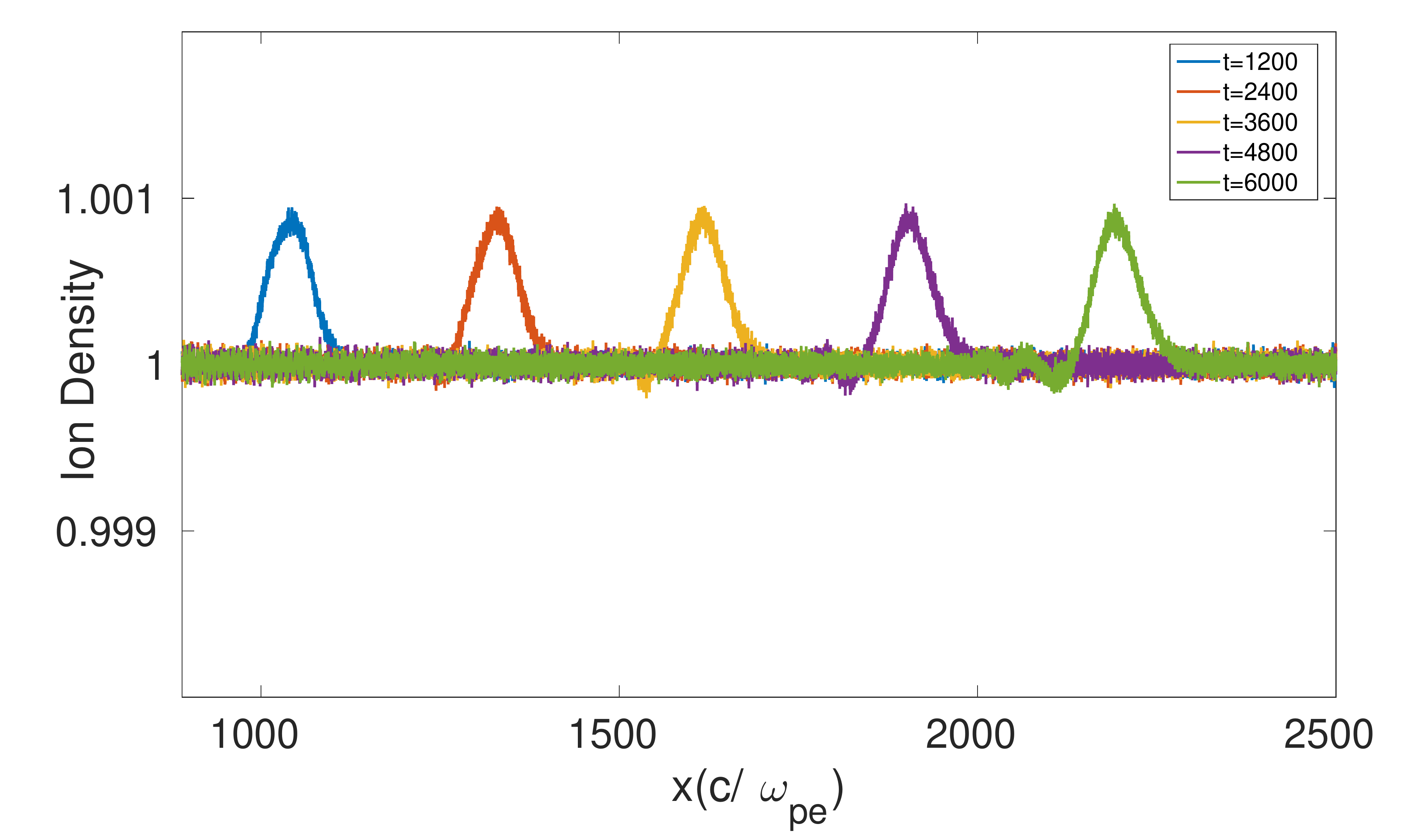}	
		\includegraphics[width=0.59\linewidth]{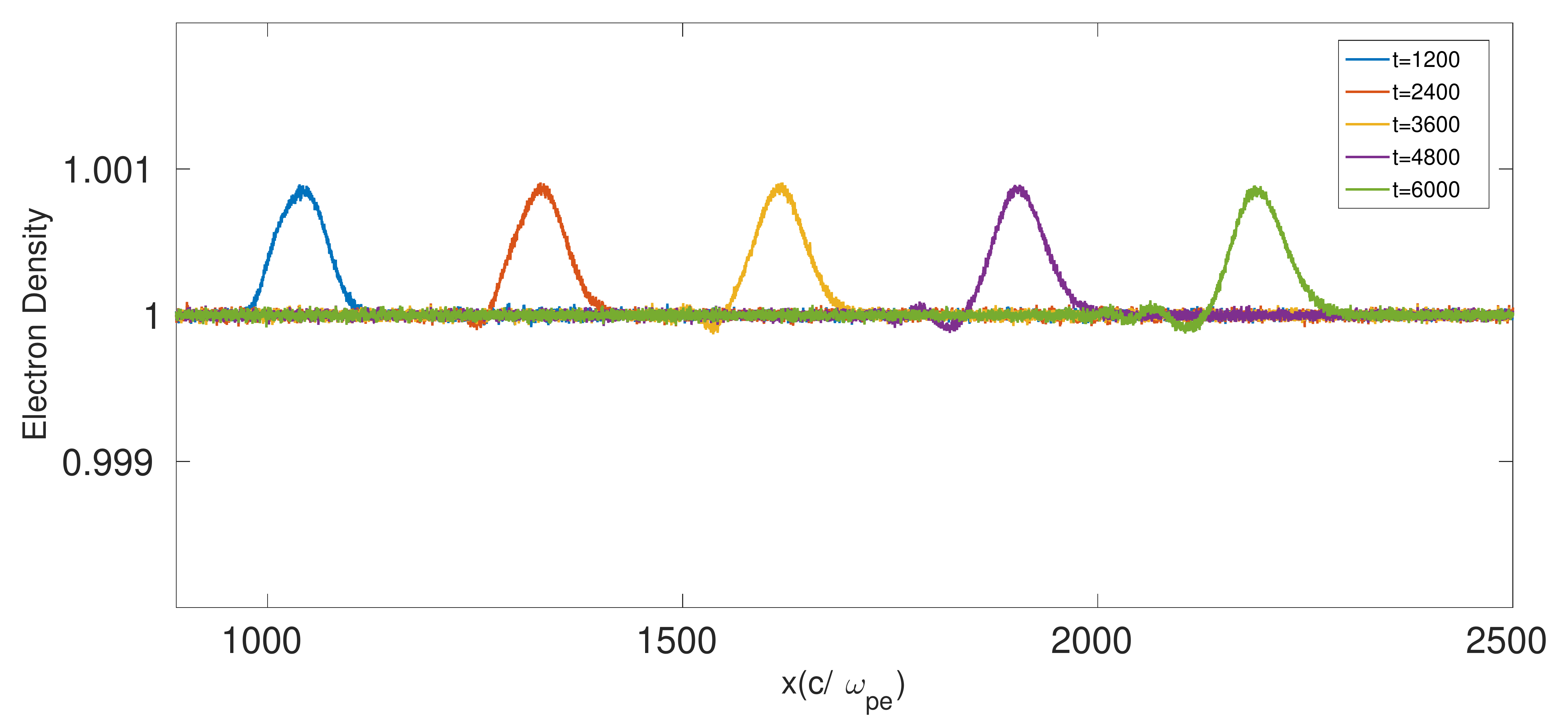}
	\caption{ Spatial variation $B_z$ , ion density and electron density of the strucutre  has been plotted at different times. Laser frequency lies in Region II for this case. The perturbation shows similar structure in $B_z$ as well as density of the charged species. Also, the structure is maintaining shape and moving with a velocity close to the Alfv\'en velocity of the medium for our choice of simulation parameters. }
	
	\label{soliton}
\end{figure*}

We now discuss the other long scale  disturbances that get generated  in the plasma by the laser, which become evident very clearly at a later time (See Fig. \ref{soliton_heating}). These disturbances have electromagnetic character as the perturbations in the $\hat{z}$ component of magnetic field has also been observed.  It can be observed from Fig. \ref{soliton} that while it is observed in the 
perturbed magnetic field $B_z$ (the applied field has been subtracted)  and $E_y$, no  formation of such a structure is observed in $E_x$.

\begin{figure*}

		\includegraphics[width=0.7\linewidth]{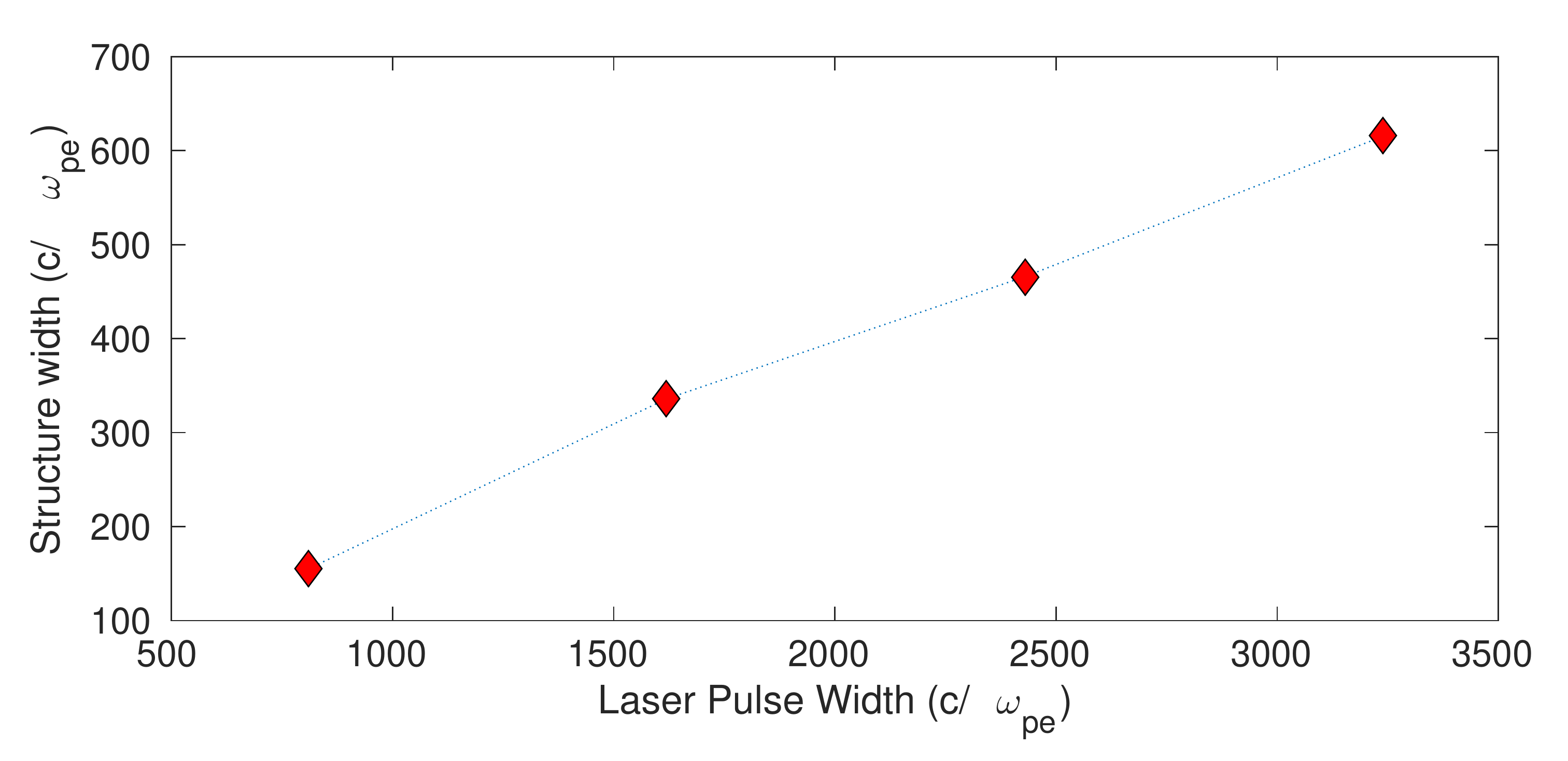}	
		\caption{ Formation of solitary peak for three different laser pulse widths lying in Region II of frequency band, keeping other parameters same. We observe that the width of the structure increases upon increasing pulse width. }
	
	\label{pulse_width}
\end{figure*}

 Furthermore, the ion and electron density perturbations both show this structure.
These structures appear as a result of the  ponderomotive force that acts on the plasma due to the 
finite longitudinal laser pulse width. The envelope of the laser pulse pushes the surface of the plasma  target, creating a magnetosonic perturbation which propagates inside the plasma.  
The frequency associated with the envelope of the laser pulse is slow so that both ions and electrons display a magnetized response and hence it excites magnetosonic perturbation. The single hump of the 
disturbance testifies that it has been excited by the  laser envelope. We 
carried out simulations with different pulse width of the laser and observe that the width of these structures  scale 
linearly with the laser pulse duration as can be  seen from  Fig. \ref{pulse_width}. We also made  
two successive laser pulse of different duration to fall on the plasma target. In this case as expected two distinct 
long scale structures get formed indicating clearly that the temporal profile of the  laser pulse is responsible 
for this. 

It is also interesting to note that since these structures are essentially excited by the envelope
of the laser pulse, they are independent of the laser frequency. Thus even when the 
laser frequency lies in the stop band of Region II, we observe the formation of these structures. 
This has been shown in Fig. \ref{soliton_heating} where we compare the formation of these long scale structures 
for both the cases of laser frequencies lying in region I and region II and having the same pulse width. It should be noted that 
in the former case the structures form along with the short scale lower hybrid (LH) excitations.  However, when the laser frequency lies in  region II,  the laser is unable to penetrate the plasma to excite the  short scale LH fluctuations are absent but the long scale 
structure continues to be present. It is yet another demonstration of the fact that the ponderomotive 
force due to the finite laser pulse induces the formation of this  long scale structures.  
The velocity of this long scale  structure is found to be  $0.24c$ which matches closely with the Alfv\'en speed ($v_A =(m_e/m_i)^{1/2} \times \omega_{ce}/ \omega_{pe}$) of the medium, which is $0.25c$ for our choice of parameters. 
 
At higher intensity of the laser field  the ponderomotive force is higher and increases the amplitude 
of these magnetosonic perturbations driving them in  nonlinear regime.  Such a disturbance  then 
forms magnetosonic solitons. 
The properties of this structure  then matches with KdV magnetosonic soliton as 
has been reported earlier \cite{kumar_soliton}.  This study can help in estimating the electromagnetic wave frequency in astrophysical observations where solitary structures and/or ion heating are observed.

\section{Conclusion}
To summarize, we have shown with the help of PIC simulations, 
the possibility of exciting electrostatic  lower hybrid perturbations in plasma with the help of a laser  in the presence of an external magnetic field.  The plasma profile 
was chosen to be sharp and the external magnetic field was chosen to be normal to both the laser 
propagation direction and the oscillatory electric field. It was shown that the necessary condition for the LH  excitation is  that the laser frequency should lie in the lower pass 
band of the $X$ mode. For this case the  electromagnetic field of the laser gets 
partially transmitted inside the plasma and then drives the lower hybrid oscillations 
by an interesting mechanism arising due to the difference between the $\vec{E} \times \vec{B}$ drift of the two species of ion and electrons in the oscillatory laser electric field and  the applied external magnetic field. We have also demonstrated  that the finite extent of the laser pulse excites long scale magnetosonic perturbations in the medium. These excitations are independent on the laser frequency and depend merely on pulse width as they get driven by the ponderomotive pressure of the laser pulse. 

Recently, there has been a lot of technological progress leading to the development of low frequency short pulse lasers such as $CO_2$ lasers and also the magnetic fields of the order of $1.2$ kilo Tesla have already been produced in the laboratory. This would,  therefore, soon open up the possibility of investigating magnetized plasma response in laser plasma related experiments, a regime  we have considered  in this paper. The importance of  LH mode is well known used in the context of magnetic confinement fusion studies where it is used for the  purpose of  current drive and heating. Here, we have shown the possibility of exciting this particular mode in the context of laser plasma interaction.  The laser experiments with magnetized response may thus 
 have important implications for several frontier experiments which rely on laser plasma interactions. For instance, it opens up a possibility of designing smarter fusion experiments which could  use the best of both inertial and magnetic confinement principles.

\section*{Acknowledgements}

The authors would like to acknowledge the OSIRIS Consortium, consisting of UCLA ans IST(Lisbon, Portugal) for providing access to the OSIRIS4.0 framework which is the  work supported by NSF ACI-1339893. AD would like to acknowledge her  J. C. Bose fellowship grant JCB/2017/000055 and the CRG/2018/000624 grant of DST for the work. The simulations for the work described in this paper were performed on Uday, an IPR Linux cluster. AV and DM would like to thank Mr. Omstavan Samant for constructive discussions and ideas.\\

\paragraph*{\bf{{References:}}}

\bibliography{LH}

\begin{figure*}
	\centering

	\includegraphics[width=0.5\linewidth]{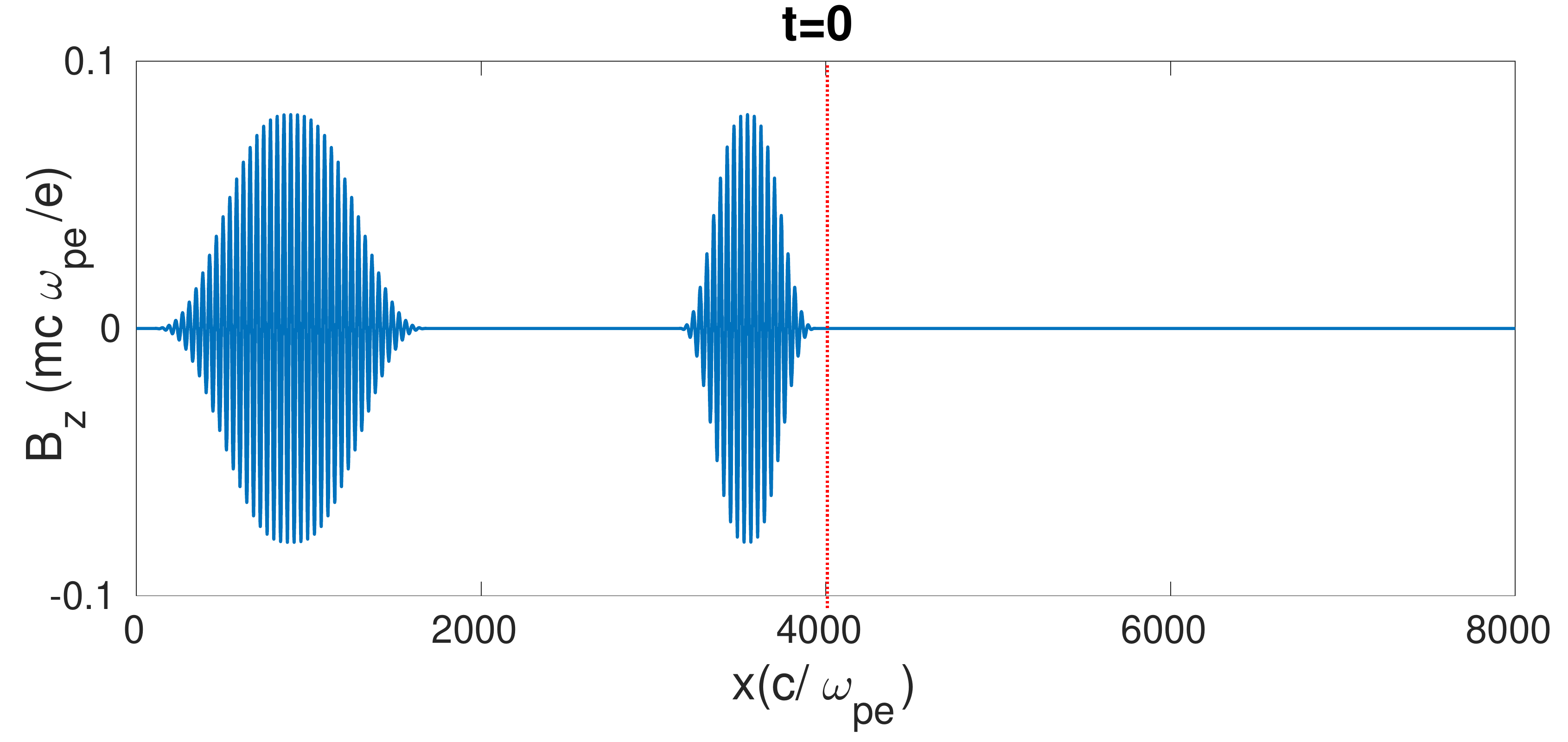}	
	\includegraphics[width=0.5\linewidth]{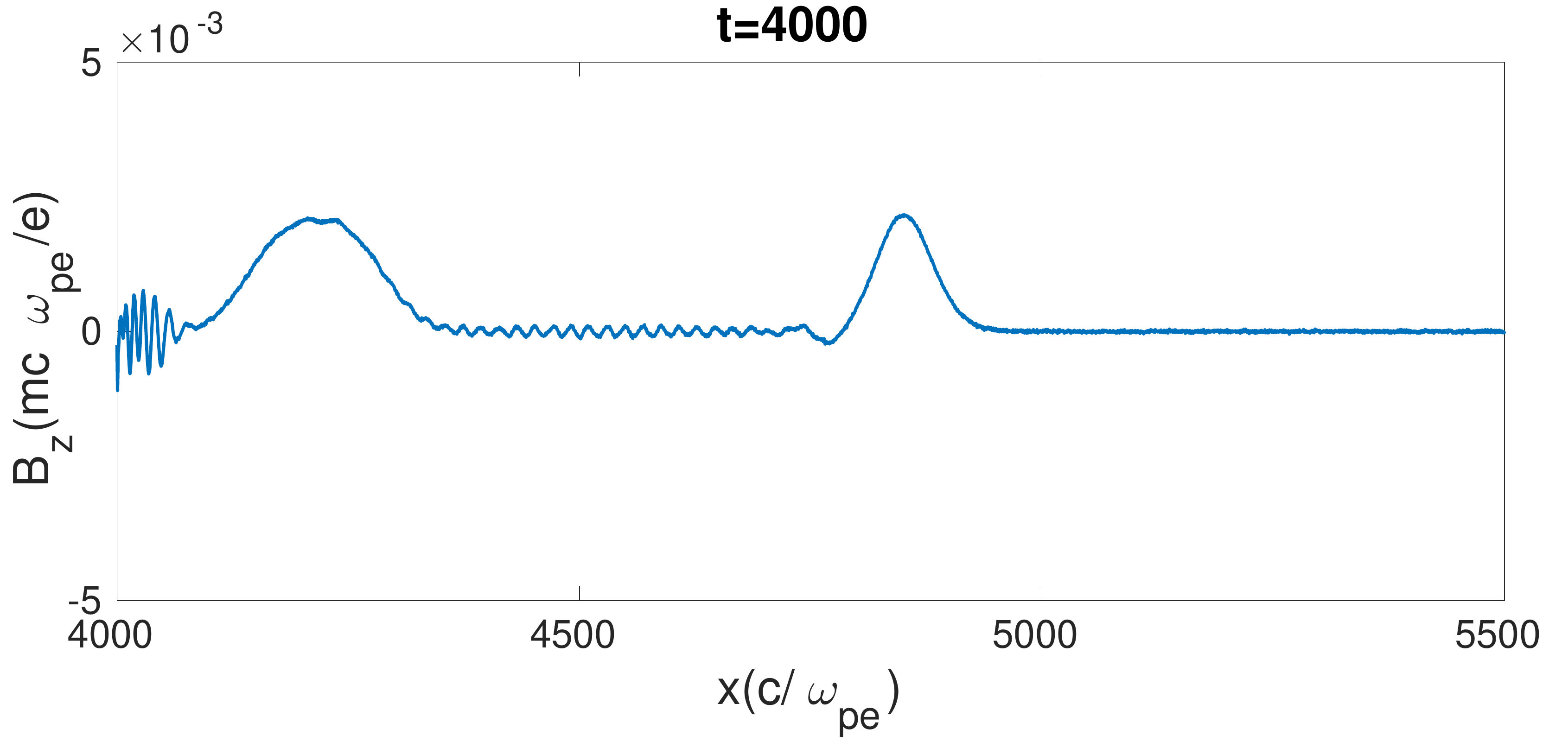}	\includegraphics[width=0.5\linewidth]{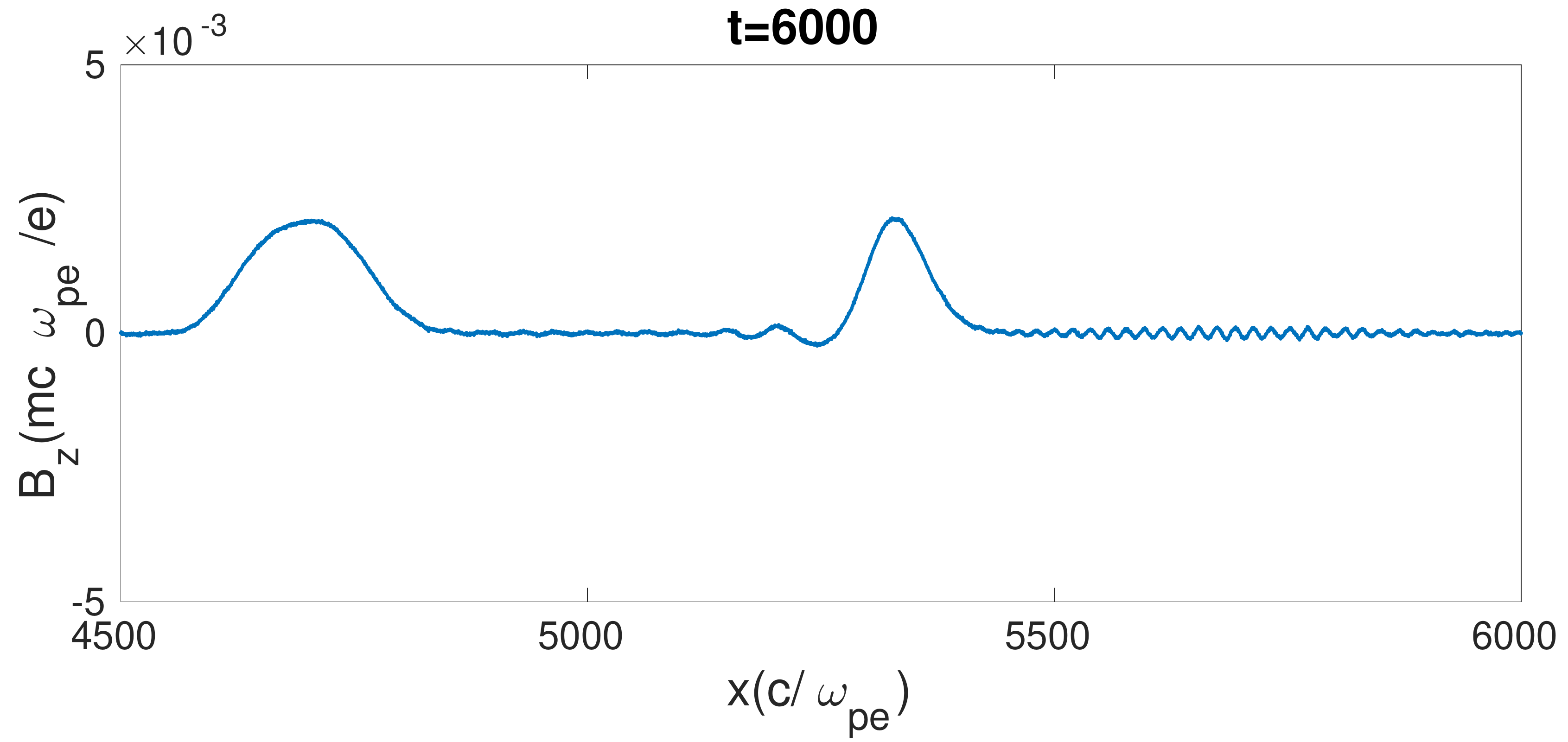}	
	
	\caption{ Spatial variation of $B_z$ at different times. At t=0, we show  two laser pulses of different widths propagating along $\hat{x}$.  The red dotted line shows plasma vaccuum interface. At subsequent times, we observe  formation of two structures of different widths as a result of interaction of the laser pulses with plasma.  Note that both the structures vary in their width depending on the initial pulse width of the laser. Both these structures move close to Alfv\'en velocity of the medium (0.25c for our chosen parameters). }
	
	\label{two_lasers}
\end{figure*}

\bibliographystyle{ieeetr}  
   
  \end{document}